\documentclass[11pt,a4paper,pdftex]{article}
\pdfoutput=1
\usepackage{jcappub}

\newcommand{\be}{\begin{eqnarray}}
\newcommand{\ee}{\end{eqnarray}}
\newcommand{\rar}{\rightarrow}

\title{Constraining the spin and the deformation parameters from the black hole shadow}

\author{Naoki Tsukamoto, Zilong Li and Cosimo Bambi$^1$ 
\note{Corresponding author}}

\affiliation{Center for Field Theory and Particle Physics \& Department of Physics,\\
Fudan University, 220 Handan Road, 200433 Shanghai, China}

\emailAdd{tsukamoto@fudan.edu.cn}
\emailAdd{zilongli@fudan.edu.cn}
\emailAdd{bambi@fudan.edu.cn}

\abstract{Within 5-10 years, very-long baseline interferometry (VLBI) facilities 
will  be able to directly image the accretion flow around SgrA$^*$, the 
super-massive black hole candidate at the center of the Galaxy, and observe 
the black hole ``shadow''. In 4-dimensional general relativity, the no-hair theorem 
asserts that uncharged black holes are described by the Kerr solution and are 
completely specified by their mass $M$ and by their spin parameter $a$. In this
paper, we explore the possibility of distinguishing Kerr and Bardeen black holes
from their shadow. In Hioki \& Maeda~(2009), under the assumption that the 
background geometry is described by the Kerr solution, the authors proposed 
an algorithm to estimate the value of $a/M$ by measuring the distortion parameter 
$\delta$, an observable quantity that characterizes the shape of the shadow. 
Here, we try to extend their approach. Since the Hioki-Maeda distortion 
parameter is degenerate with respect to the spin and possible deviations from 
the Kerr solution, one has to measure another quantity to test the Kerr black 
hole hypothesis. We study a few possibilities. We find that it is extremely difficult 
to distinguish Kerr and Bardeen black holes from the sole observation of the 
shadow, and out of reach for the near future. The combination of the measurement 
of the shadow with possible accurate radio observations of a pulsar in a compact 
orbit around SgrA$^*$ could be a more promising strategy to verify the Kerr black 
hole paradigm.}


\begin{document}

\maketitle


\section{Introduction}

In 4-dimensional general relativity, the no-hair theorem guarantees that uncharged
black holes (BHs) are only described by the Kerr solution, which is completely
specified by two parameters; that is, the BH mass $M$ and the BH spin angular 
momentum $J$~\cite{hair1,hair2,hair3}. A fundamental limit for a Kerr 
BH is the bound $|a| \le M$, where $a = J/M$ is the BH spin 
parameter\footnote{Throughout the paper, we use units in which $G_{\rm N} = c = 1$, 
unless stated otherwise.}. This is just the condition for the existence of the event 
horizon. Astrophysical BH candidates are stellar-mass compact objects in X-ray 
binary systems and super-massive bodies at the center of every normal 
galaxy~\cite{narayan}. They are thought to be the Kerr BHs of general relativity 
simply because they cannot be explained otherwise without introducing new physics, 
but there is no evidence that the spacetime around them is described by the Kerr 
metric.

In the last decade, there have been both significant theoretical work to understand
the accretion process onto BH candidates and new observational facilities, so 
that it might be possible to test the actual nature of these objects in a near 
future from the properties of the electromagnetic radiation emitted by the accreting
material~\cite{rev1,rev2}. A large number of methods have been proposed, including
the study of the thermal spectrum of thin accretion disks~\cite{cfm1,cfm2,cfm3,cfm4}, 
the analysis of the K$\alpha$ iron line~\cite{iron1,iron2,iron3}, the observation of the
so-called quasi-periodic oscillations (QPOs)~\cite{qpo1,qpo2,qpo3,qpo4}, the
measurement of the radiative efficiency~\cite{eta1,eta2,eta3,eta4}, and the estimate 
of the jet power~\cite{jet1,jet2,jet3}. The main difficulty to achieve this goal is that
the properties of the electromagnetic radiation from a Kerr BH with dimensionless
spin parameter $a/M$ can be very similar, and practically indistinguishable, from the 
ones of non-Kerr objects with different spin. In other words, it is usually impossible
to constrain at the same time the value of the spin parameter and possible deviations
from the Kerr solution. For some metrics, the combination of the analysis of the disk's 
thermal spectrum and of the K$\alpha$ iron line cannot break the degeneracy between 
the spin and the deformation parameters, while in other cases that can be achieved 
only with very good measurements~\cite{cfm-iron}. The combination of the analysis
of the thermal spectrum of thin disks and the estimate of the jet power can potentially
do the job~\cite{jet2,jet3}, but the latter is not yet a mature technique and therefore 
it cannot yet be used to test fundamental physics. The result is that right now we can 
only rule out the possibility that BH candidates are some kinds of very exotic objects,
like some types of wormholes~\cite{wh} or some exotic compact objects without 
event horizon~\cite{bs,bs2}. The non-observation of electromagnetic radiation emitted 
by the possible surface of these objects may also be interpreted as an indication 
for the existence of an event horizon~\cite{hor1,hor2} (but see~\cite{hor3,hor4}). 
However, more reasonable alternatives, like non-Kerr BHs, are difficult to test.

Recent observations of SgrA$^*$ at mm wavelength suggest that, hopefully within 
about 5~years, very-long baseline interferometry (VLBI) facilities at mm/sub-mm
wavelength will be able to directly image the accretion flow around the super-massive 
BH candidate at the center of our Galaxy with a resolution of the order its 
gravitational radius $r_g = M$~\cite{vlbi1,vlbi2}. These observations will open a
completely new window to test gravity in the strong field regime and, in particular, 
to verify if SgrA$^*$ is a Kerr BH, as expected from general relativity. The main
goal of these experiments is the observation of the ``shadow'' of SgrA$^*$. The
shadow of a BH is a dark area over a brighter background observed by directly
imaging the accretion flow around the compact object~\cite{sh1,sh2}. While the
intensity map of the image depends on complicated astrophysical processes
related to the accretion properties and the emission mechanisms, the exact 
shape of the shadow is only determined by the background geometry, being the
apparent photon capture sphere as seen by a distant observer. A very accurate 
detection of the boundary of the shadow can thus provide information on the
geometry around SgrA$^*$ and test the Kerr BH hypothesis. Starting from 
Refs.~\cite{sh3,sh4}, a number of tests has been proposed in the literature and 
shadows in different background metrics have been calculated by different 
groups~\cite{sh5,sh6,sh7,sh8,sh9,sh10,sh11,sh12}. For a recent review, see
e.g. Ref.~\cite{rev-sh}.

At first approximation, the shape of the shadow is a circle. The radius of the 
circle corresponds to the apparent photon capture radius, which, for a given
metric, is set by the mass of the compact object and its distance from us. For
SgrA$^*$ and for the other nearby super-massive BH candidates, these two 
quantities are currently not known with good precision, and therefore the 
observation of the size of the shadow may not be used to test the spacetime 
geometry around the compact object (but see Ref.~\cite{sh11} and the 
conclusions of the present work). The shape of the shadow is usually thought 
to be the key-point. The first order correction to the circle is due to the spin, 
as the photon capture radius is different for co-rotating and counter-rotating 
particles. The boundary of the shadow has thus a dent on one side: the 
deformation is more pronounced for an observer on the equatorial plane 
(viewing angle $i = 90^\circ$) and decreases as the observer moves towards 
the spin axis, to completely disappear when $i = 0^\circ$ or $180^\circ$. 
Possible deviations from the Kerr solutions usually introduce smaller 
corrections and therefore they can be detected only in the case of excellent 
data.

In Ref.~\cite{z3}, two of us have studied the measurement of the Kerr spin 
parameter of Kerr BHs and non-Kerr regular BHs; that is, we measured the 
spin parameter $a/M$ from the shape of the shadow of a BH assuming it was 
a Kerr BH. We used the procedure proposed in Ref.~\cite{maeda}, which 
is based on the determination of the distortion parameter $\delta = D/R$, 
where $D$ and $R$ are, respectively, the dent and the radius of the 
shadow. In the case of non-Kerr BHs, this technique provides the correct value 
of $a/M$ for non-rotating objects, but a quite different spin for near extremal 
states. If we compare this measurement with the frequency of the innermost
stable circular orbit (ISCO) that can be potentially obtained by the observations
of blobs of plasma orbiting around the BH candidate~\cite{z4}, we find that 
the nature of the object may be tested in the case of a non-rotating 
or slow-rotating BHs, while that seems to be impossible for near extremal 
states, as the two techniques essentially provide the same information on the 
spacetime geometry.

In the present paper, we consider a different approach to test the Kerr geometry
around SgrA$^*$. We assume to have good observational data of the BH shadow 
and we try to measure two parameters. One parameter of the shadow can indeed
only determine one parameter of the background geometry, which is enough in 
the case of the Kerr metric where there is only the spin. If we want to test the Kerr 
nature of the BH candidate, the spacetime metric will be also characterized by 
one (or more) deformation parameter(s), measuring possible deviations from the Kerr 
solution. In general, the Hioki-Maeda distortion parameter $\delta$ is degenerate with 
respect to the spin and the deformation parameters, in the sense that the same 
value of $\delta$ is found for any deformation parameter for a particular value of 
$a/M$. With the measurement of another parameter of the shadow, it is 
possible to break such a degeneracy and eventually test the Kerr metric of the 
spacetime. We explore three possibilities. We introduce a second distortion 
parameter, $\epsilon$, which characterizes possible deviations from the shape of 
the shadow of a Kerr BH. While a similar approach may sound the most natural 
extension of Ref.~\cite{maeda}, it turns out that Kerr BHs and non-Kerr regular 
BHs have very similar shapes and such a small difference is very difficult to 
detect in true observational data. We thus consider the possibility of measuring 
the off-set of the center of the shadow with respect to the actual position of the 
BH. Even in this case, the approach can potentially distinguish Kerr BHs and 
non-Kerr regular BHs, but an accurate measurement of the BH position is very 
challenging, at least now. The third and last case is the measurement of the radius 
of the shadow, $R$. Here, we need good measurements of the BH mass and distance 
from us, which are definitively not available today, but they could be possible 
in the future, for instance from accurate radio observations of a pulsar orbiting
SgrA$^*$ with a period shorter than 1 year. The same pulsar could also provide
a precise estimate of the spin parameter (obtained in the weak field), to be
compared with the Hioki-Maeda distortion parameter $\delta$ of the strong
gravity regime. It seems therefore that the sole observation of the shadow
cannot distinguish Kerr and Bardeen BHs, while the combination of the shadow
and pulsar observations is more promising to probe the geometry around SgrA$^*$.

The content of the paper is as follows. In Section~\ref{s-sh}, we review the calculation
of the BH shadow, while in Section~\ref{s-spin} we review the
procedure proposed in Ref.~\cite{maeda} to infer the spin parameter from the 
determination of the distortion parameter $\delta$. In Section~\ref{s-def}, we
explore the possibility of testing the Kerr metric from the combination of the 
estimates of the Hioki-Maeda distortion parameter $\delta$ with, respectively, 
the distortion parameter $\epsilon$, the position of the center of the shadow 
with respect to the one of the BH, and the shadow radius $R$. Section~\ref{s-d} is 
devoted to the discussion. Summary and conclusions are in Section~\ref{s-c}.

\section{Black hole shadow \label{s-sh}}

If a BH is surrounded by an optically thin and geometrically thick accretion flow,
a distant observer sees a dark area over a
brighter background. Such a dark area is the ``shadow'' of the BH. The boundary 
of the shadow corresponds to the apparent image of the photon capture sphere 
and therefore it only depends on the geometry of the background~\cite{sh1,sh2}. 
In this section, we briefly review the calculation of the BH shadow. In the case of
the Kerr metric, the line element in Boyer-Lindquist coordinates is
\be\label{eq-metric}
ds^2 &=& - \left(1 - \frac{2 M r}{\Sigma}\right) dt^2 
- \frac{4 a M r \sin^2 \theta}{\Sigma} dt d\phi
+ \frac{\Sigma}{\Delta} dr^2 + \Sigma d\theta^2 \nonumber\\ 
&& + \left(r^2 + a^2 + \frac{2 a^2 M r \sin^2\theta}{\Sigma}\right)
\sin^2\theta d\phi^2 \, ,
\ee
where
\be\label{eq-metric2}
\Sigma = r^2 + a^2 \cos^2\theta \, , \quad
\Delta = r^2 - 2 M r + a^2 \, ,
\ee
$M$ is the BH mass, and $a = J/M$. The photon motion is governed by the 
equations~\cite{sh1}
\be
\Sigma \bigg(\frac{d t}{d \lambda}\bigg)
&=& \frac{A E - 2 a M r L_z}{\Delta} \, , \\
\Sigma^2\bigg(\frac{d r}{d \lambda}\bigg)^2 
&=& \mathcal{R} \, ,   \label{eq-radial} \\ 
\Sigma^2\bigg(\frac{d \theta}{d \lambda}\bigg)^2 
&=& \Theta \, , \\
\Sigma \bigg(\frac{d \phi}{d \lambda}\bigg)
&=& \frac{2 a M r E + \left(\Sigma - 2 M r\right) L_z \csc^2 \theta}{\Delta} \, ,
\ee
where $\lambda$ is an affine parameter, and
\be
\mathcal{R} &=& E^2 r^4+\left(a^2E^2-L_z^2-\mathcal{Q}\right)r^2 
+2 M \left[(aE-L_z)^2+\mathcal{Q}\right]r-a^2\mathcal{Q}\, ,   \label{eq-R} \\
\Theta &=& \mathcal{Q} \left(a^2 E^2 - L_z^2 \csc^2 \theta \right)
\cos^2 \theta \, , \\
A &=& \left(r^2 + a^2\right)^2 - a^2 \Delta \sin^2 \theta \, .
\ee
$E$ and $L_z$ are, respectively, the conserved photon energy and the conserved 
component of the photon angular momentum parallel to the BH spin. $\mathcal{Q}$
is the Carter constant
\be
\mathcal{Q} &=& p_\theta^2+\cos^2\theta
\bigg(\frac{L_z^2}{\sin^2\theta}-a^2E^2\bigg) \, ,
\ee
and $p_\theta = \Sigma \frac{d\theta}{d\lambda}$ is the canonical momentum 
conjugate to $\theta$.

Motion is only possible when $\mathcal{R}(r) \geq 0$, and therefore the analysis of 
the position of the roots of $\mathcal{R}(r)$ can be used to distinguish the capture
from the scattered orbits. The three kinds of photon orbits are:
\begin{enumerate}
\item {\it Capture orbits}: $\mathcal{R}(r)$ has no roots for $r \geq r_+$, where 
$r_+$ is the radial coordinate of the BH event horizon. In this case, photons come 
from infinity and then cross the horizon.
\item {\it Scattering orbits}: $\mathcal{R}(r)$ has real roots for $r \geq r_+$, which
correspond to the photon turning points. If the photons come from infinity, they 
reach a minimum distance from the BH, and then go back to infinity. 
\item {\it Unstable orbits of constant radius}: these orbits separate the capture and 
the scattering orbits and are determined by
\be
\mathcal{R}(r_*) = \frac{\partial \mathcal{R}}{\partial r}(r_*)=0 \, , \quad
{\rm and} \quad
\frac{\partial^2 \mathcal{R}}{\partial r^2}(r_*) \geq 0 \, ,   \label{unstable}
\ee
where $r_*$ is the larger real root of $\mathcal{R}$.
\end{enumerate}
The boundary of the shadow of a BH can be determined by finding the unstable
orbits of constant radius.

Since the photon trajectories are independent of the photon wavelength, it is convenient
to introduce the parameters $\xi = L_z/E$ and $\eta = \mathcal{Q}/E^2$. $\xi$ and 
$\eta$ are related to the ``celestial coordinates'' $\alpha$ and $\beta$ of the image
plane of the distant observer by
\be
\alpha = \frac{\xi}{\sin i}\, ,  \quad
\beta = \pm (\eta+a^2\cos^2 i-\xi^2\cot^2 i)^{1/2}\, ,
\ee
where $i$ is the angular coordinate of the observer at infinity. Every photon orbit
can be characterized by the constants of motion $\xi$ and $\eta$.  The boundary
of the BH shadow is represented by a closed curve determined by the set of unstable 
circular orbits ($\xi_c$, $\eta_c$) on the plane of the distant observer. From 
Eqs.~(\ref{eq-R}) and (\ref{unstable}), the equations determining the unstable 
orbits of constant radius are
\be
\mathcal{R} &=& r^4+(a^2-\xi_c^2-\eta_c)r^2+2M[\eta_c+(\xi_c-a)^2]r  
-a^2\eta_c = 0 \, ,  \nonumber \\
\frac{\partial \mathcal{R}}{\partial r} &=& 4r^3+2(a^2-\xi_c^2-\eta_c)r
+2M[\eta_c+(\xi_c-a)^2]  = 0 \, .    \label{eq-critical}
\ee
In the Schwarzschild background ($a=0$), the BH shadow is a circle of radius
\be
R = 3 \sqrt{3} \, M \approx 5.196 \, M \, .
\ee
If $a\neq 0$, one finds
\be\label{xieta}
\xi_c &=& \frac{M(r_*^2-a^2)-r_*(r_*^2-2Mr_*+a^2)}{a(r_*-M)} \, , \nonumber \\
\eta_c &=& \frac{r_*^3 [4a^2M-r_*(r_*-3M)^2]}{a^2(r_*-M)^2} \, ,
\ee
where $r_*$ is the radius of the unstable orbit. The shadow of Kerr BHs can be
found in many papers in the literature~\cite{maeda}.

In the present work, we want to figure out how we can distinguish Kerr
and Bardeen BHs from the observation of the BH shadow. In Boyer-Lindquist
coordinates, the line element of the rotating Bardeen metric has the same
form as the Kerr one, Eqs.~(\ref{eq-metric}) and (\ref{eq-metric2}), with the 
mass $M$ replaced by $m$~\cite{reg1,reg2}:
\be
M \rar m = M \left(\frac{r^2}{r^2 + g^2}\right)^{3/2} \, ,  \label{m-bardeen}
\ee
without changing $a$ (at least in the simplest form, see Ref.~\cite{reg2} for more 
details). Here $g$ can be interpreted as the magnetic charge of a 
non-linear electromagnetic field\footnote{As found in Ref.~\cite{reg1}, the 
Bardeen metric can be obtained as an exact solution of Einstein's equations 
coupled to a non-linear electromagnetic field. The latter allows to avoid
the no-hair theorem.} or just as a quantity introducing a deviation from 
the Kerr solution. The position of the even horizon is still given by the largest
root of $\Delta = 0$ and there is a bound on the maximum value of the spin 
parameter, above which there are no BHs. The maximum value of $a/M$ is 
1 for $g/M = 0$ (Kerr case), and decreases as $g/M$ increases, to reach 0
for $g/M = \sqrt{16/27} \approx 0.7698$. There are no BHs for $g/M > \sqrt{16/27}$. 
The situation is similar to the Kerr-Newman solution, where the maximum value 
of $a/M$ is 1 for a vanishing electric charge $Q$ (Kerr case), and decreases 
as $Q$ increases, to reach 0 for $Q/M = 1$. The Bardeen metric can be seen 
as the prototype of a large class of metrics, in which the line element in Boyer-Lindquist 
coordinates has the same form as the Kerr one, with $M$ replaced by a
function $m(r)$ that depends only on the radial coordinate. Such a family
of metrics includes the Kerr-Newman solution, in which $m = M - Q^2/2r$.

Since the Bardeen metric (as well as all the other metrics in this family) has the
same nice properties as the Kerr one, in particular there exists the Carter 
constant $\mathcal{Q}$ and the equations of motion are separable in 
Boyer-Lindquist coordinates, it is straightforward to generalize the above 
calculations of the shadow of a Kerr BH to the Bardeen case. The system in
Eq.~(\ref{eq-critical}) is replaced by
\be
\mathcal{R} &=& r^4+(a^2-\xi_c^2-\eta_c)r^2+2m[\eta_c+(\xi_c-a)^2]r  
-a^2\eta_c = 0 \, ,  \nonumber \\
\frac{\partial \mathcal{R}}{\partial r} &=& 4r^3+2(a^2-\xi_c^2-\eta_c)r
+2m[\eta_c+(\xi_c-a)^2] f = 0 \, , 
\ee
with $m$ given by Eq.~(\ref{m-bardeen}) and $f$ defined by 
\be
f = 1 + \frac{r}{m} \frac{dm}{dr} = \frac{r^2 + 4 g^2}{r^2 + g^2} \, .
\ee
The counterpart of Eq.~(\ref{xieta}) is
\be\label{eq-fff}
\xi_c &=& \frac{m[(2 - f)r_*^2-fa^2]-r_*(r_*^2-2mr_*+a^2)}{a(r_*-fm)} \, , \nonumber \\
\eta_c &=& \frac{r_*^3 \{4(2-f)a^2m-r_*[r_*-(4 - f)m]^2\}}{a^2(r_*- fm)^2} \, ,
\ee
with $m = m(r_*)$ and $f=f(r_*)$.
Example of shadows of Bardeen BHs are reported in Ref.~\cite{z3}.
Let us note that Eq.~(\ref{eq-fff}) does not hold only for the Bardeen metric, 
but in the large class of BH solutions in which the line element is given by 
Eqs.~(\ref{eq-metric}) and (\ref{eq-metric2}) with $M$ replaced by some $m(r)$
that depend on the radial coordinate only.

\begin{figure}
\begin{center}
\includegraphics[type=pdf,ext=.pdf,read=.pdf,width=8.7cm]{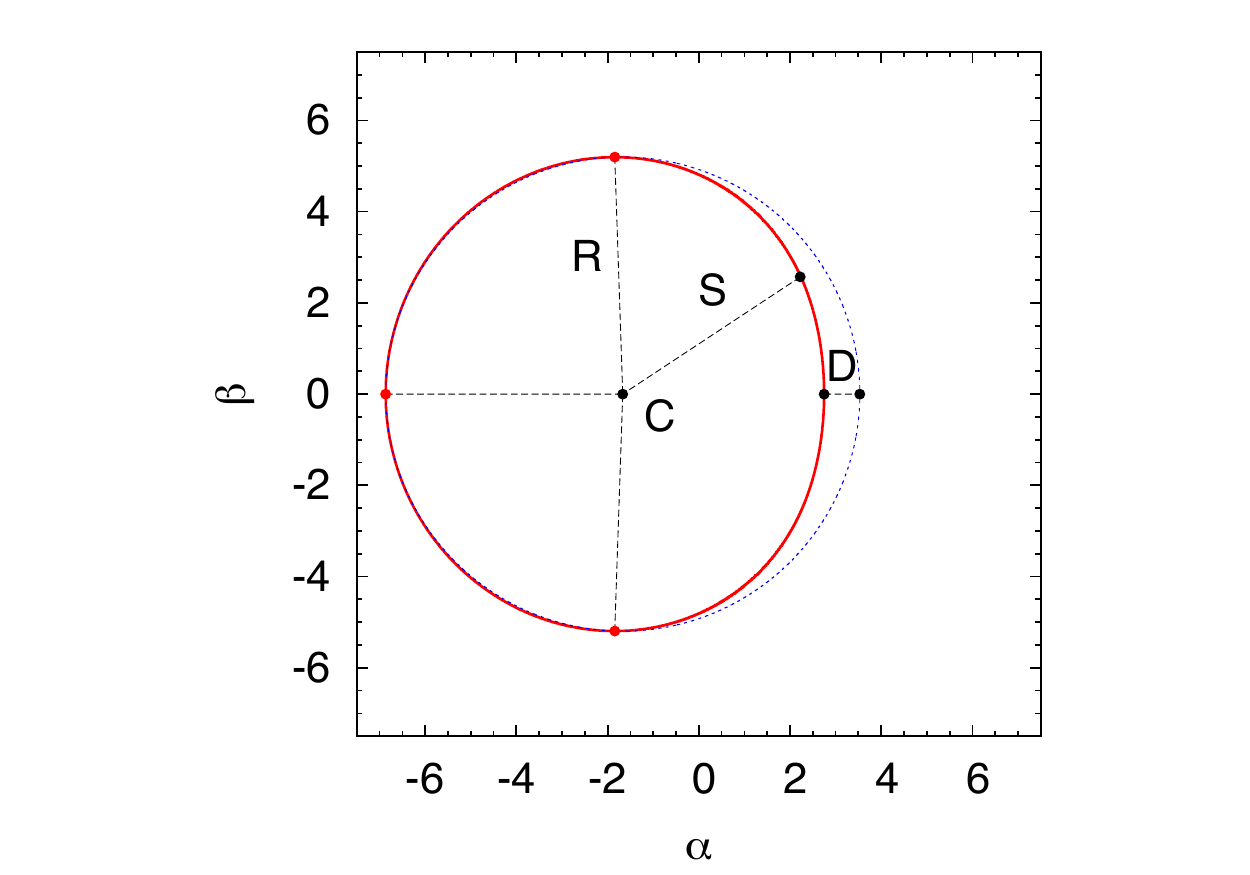}
\end{center}
\vspace{-0.5cm}
\caption{BH shadow with the three parameters that approximately characterize 
its shape: the radius $R$, the dent $D$, and the distance $S$. $R$ is defined as the 
radius of the circle passing through the three red points, located at the top 
($\beta = \beta_{\rm max}$), bottom, 
and most left end of the shadow. $D$ is the difference between the most right points 
of the circle and of the shadow. $S$ is the distance between the center of the circle, 
$C$, and the most right end of the shadow at $\beta = \beta_{\rm max}/2$. The 
Hioki-Maeda distortion parameter is $\delta = D/R$. The second distortion parameter 
is $\epsilon = S/R$. $\alpha$ and $\beta$ in units $M=1$. See the text for more details.}
\label{fig1}
\end{figure}

\section{Measuring the Kerr spin parameter from the observation of the shadow \label{s-spin}}

The boundary of the BH shadow corresponds to the apparent image of the photon 
capture sphere as seen by the distant observer and it is only determined by the 
background geometry. If we want to infer the value of the parameters of the metric, 
it is convenient to figure out the features of the shadow that better characterize its 
shape and that can be measured from the shadow image. This strategy was first 
proposed in~\cite{maeda} to estimate the value of the spin parameter $a/M$ from the 
observation of the shadow of a Kerr BH. In this section, we will briefly review their 
approach, while in the next section it will be extended to test the Kerr metric of 
BH candidates.

At first approximation, the boundary of the shadow is a circle, and it is exactly a 
circle in the case of a static spherically symmetric solution (like the Schwarzschild 
metric) or in the one of a stationary axisymmetric solution in which the distant 
observer is located along the symmetry axis (like the Kerr metric and a viewing 
angle $i=0^\circ$ or $180^\circ$). We can thus approximate the shadow with a 
circle passing through the three points located, respectively, at the top position, 
bottom position, and most left end of its boundary (the three red points in Fig.~\ref{fig1}). 
The radius of the shadow, $R$, is defined as the radius of this circle.

The first order correction to the circle is due to the spin, because of the spin-orbit 
coupling between the photon and the BH. The gravitational force is indeed 
stronger if the photon angular momentum is antiparallel to the BH spin (the 
photon capture radius is thus larger), and weaker in the opposite case (the 
photon capture radius is smaller). The result is that the shadow has a dent 
on one side, which is larger for a viewing angle $i = 90^\circ$, and reduces as 
the distant observer move to the axis of symmetry, to completely disappear 
when $i=0^\circ$ or $180^\circ$. We define the dent $D$ as the distance between
the right endpoints of the circle and of the shadow, see Fig.~\ref{fig1}. We can
then introduce the Hioki-Maeda distortion parameter $\delta = D/R$, which
is a quantity that can be measured from the image of the shadow and can 
be used to characterize its shape~\cite{maeda}.

\begin{figure}
\begin{center}
\hspace{-1cm}
\includegraphics[type=pdf,ext=.pdf,read=.pdf,width=8.7cm]{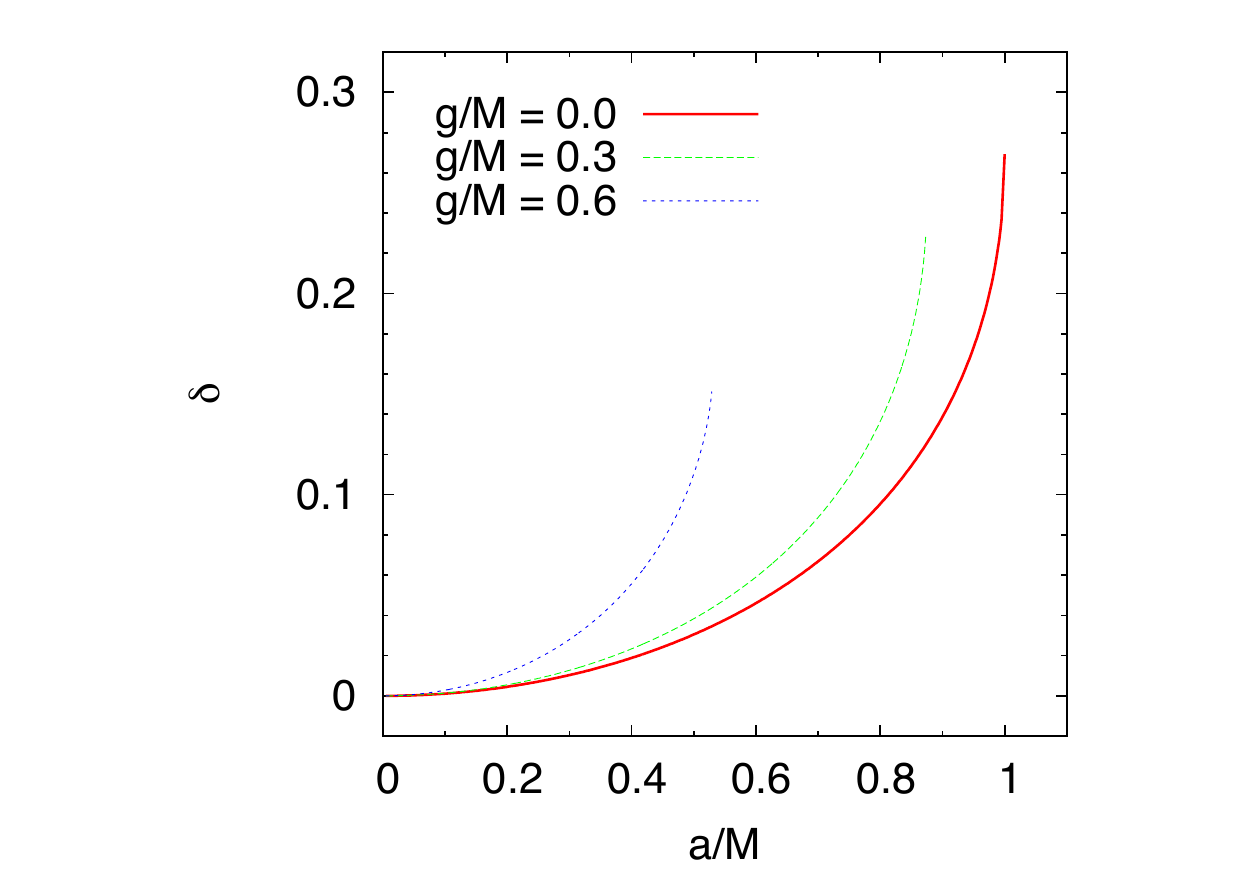}
\hspace{-1.5cm}
\includegraphics[type=pdf,ext=.pdf,read=.pdf,width=8.7cm]{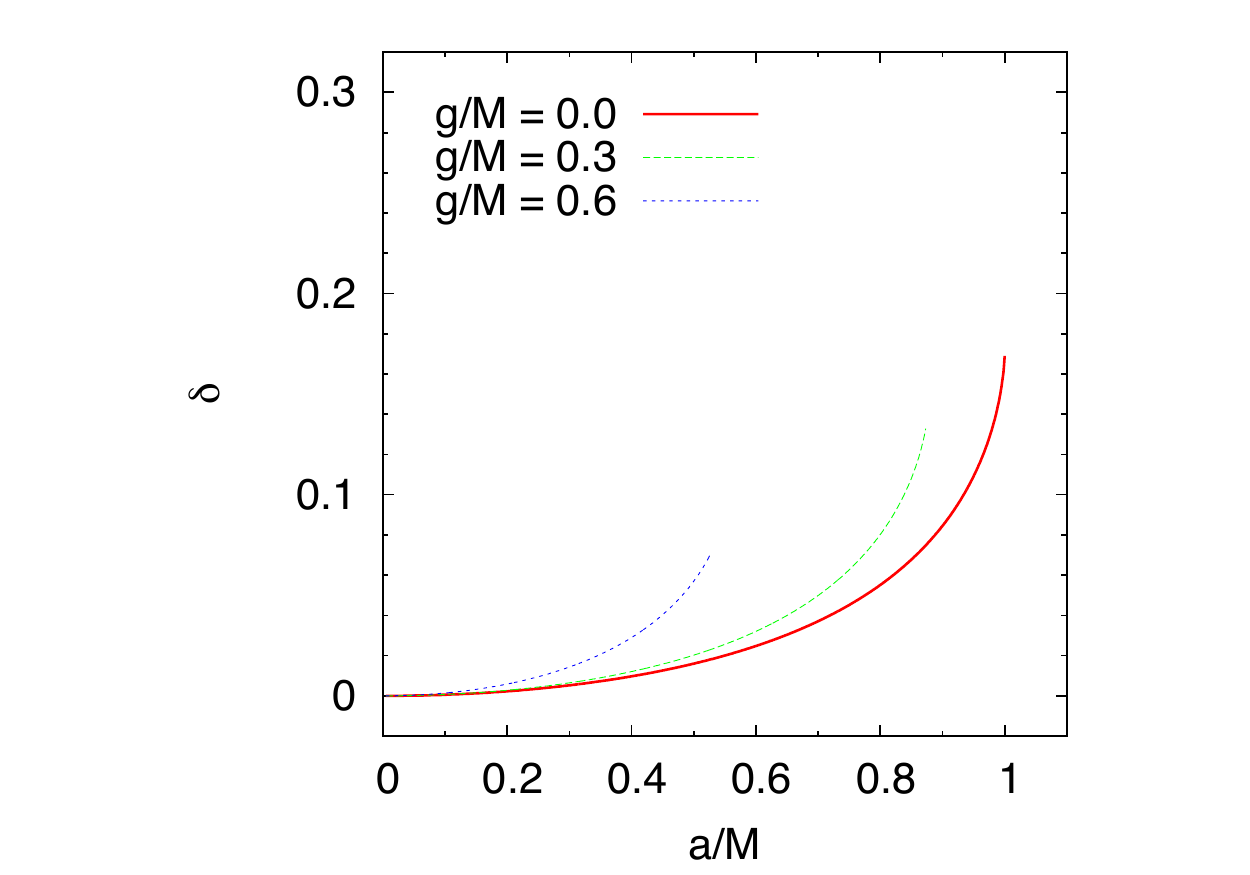}
\end{center}
\vspace{-0.5cm}
\caption{Hioki-Maeda distortion parameter $\delta$ as a function of the spin
parameter $a/M$ for Kerr BHs (red solid line), Bardeen BHs with $g/M=0.3$
(green dashed line), and Bardeen BHs with $g/M=0.6$ (blue dotted line).
The inclination angle is $i = 90^\circ$ (left panel) and $45^\circ$ (right panel).
The maximum value of the Hioki-Maeda distortion parameter is 
$\delta = 7/26 \approx 0.269$ in the case of an extremal Kerr BH ($a/M = 1$) 
and a viewing angle $i = 90^\circ$.}
\label{fig2}
\end{figure}

If we know that the spacetime geometry is described by the Kerr solution and 
we have an independent estimate of the viewing angle $i$, the measurement 
of the distortion parameter provides an estimate of the BH spin parameter $a/M$.
Indeed, for a give $i$, there is a one-to-one correspondence between $a/M$ 
and $\delta$, and the function $\delta(a/M)$ can be inverted to obtain $a/M|^{\rm Kerr}(\delta)$.
If we relax the Kerr BH hypothesis and we want to test the nature of the BH candidate, 
we have to introduce at least one parameter that quantifies possible deviations 
from the Kerr geometry. If we adopt the Bardeen metric, this role is played by the 
Bardeen charge $g$. Now the boundary of the shadow, as well as all the other 
properties of the background metric at small radii, depends on both $a/M$ and $g/M$. 
The distortion parameter is $\delta(a/M,g/M)$ and it is not possible to infer $a/M$ 
without an independent estimate of $g/M$. If we assume that the object is a Kerr 
BH even if $g/M \neq 0$, we can determine the so-called Kerr spin parameter
\be
a/M|^{\rm Kerr} = a/M|^{\rm Kerr} [ \delta(a/M,g/M) ] \, ,
\ee
which provides a wrong value of the spin for $a/M \neq 0$~\cite{z3}. Fig.~\ref{fig2}
shows the Hioki-Maeda distortion parameter $\delta$ as a function of the BH
spin parameter $a/M$ for the case of Kerr BHs and Bardeen BHs with $g/M = 0.3$
and $0.6$. It is clear that the same distortion parameter $\delta$ can characterize
the shadow of a Kerr BH with spin $a/M$ or of a Bardeen BH with lower spin.

\begin{figure}
\begin{center}
\hspace{-1cm}
\includegraphics[type=pdf,ext=.pdf,read=.pdf,width=8.7cm]{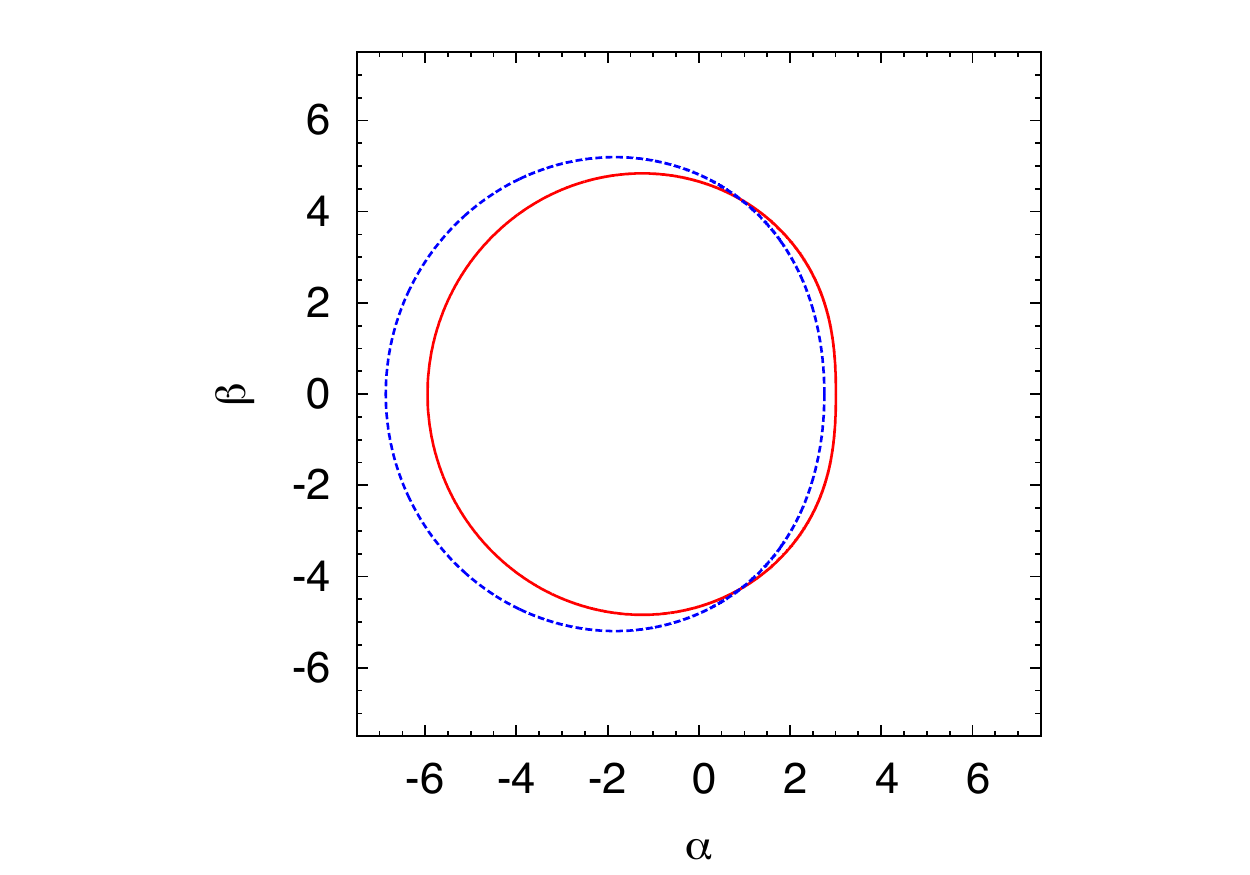}
\hspace{-1.5cm}
\includegraphics[type=pdf,ext=.pdf,read=.pdf,width=8.7cm]{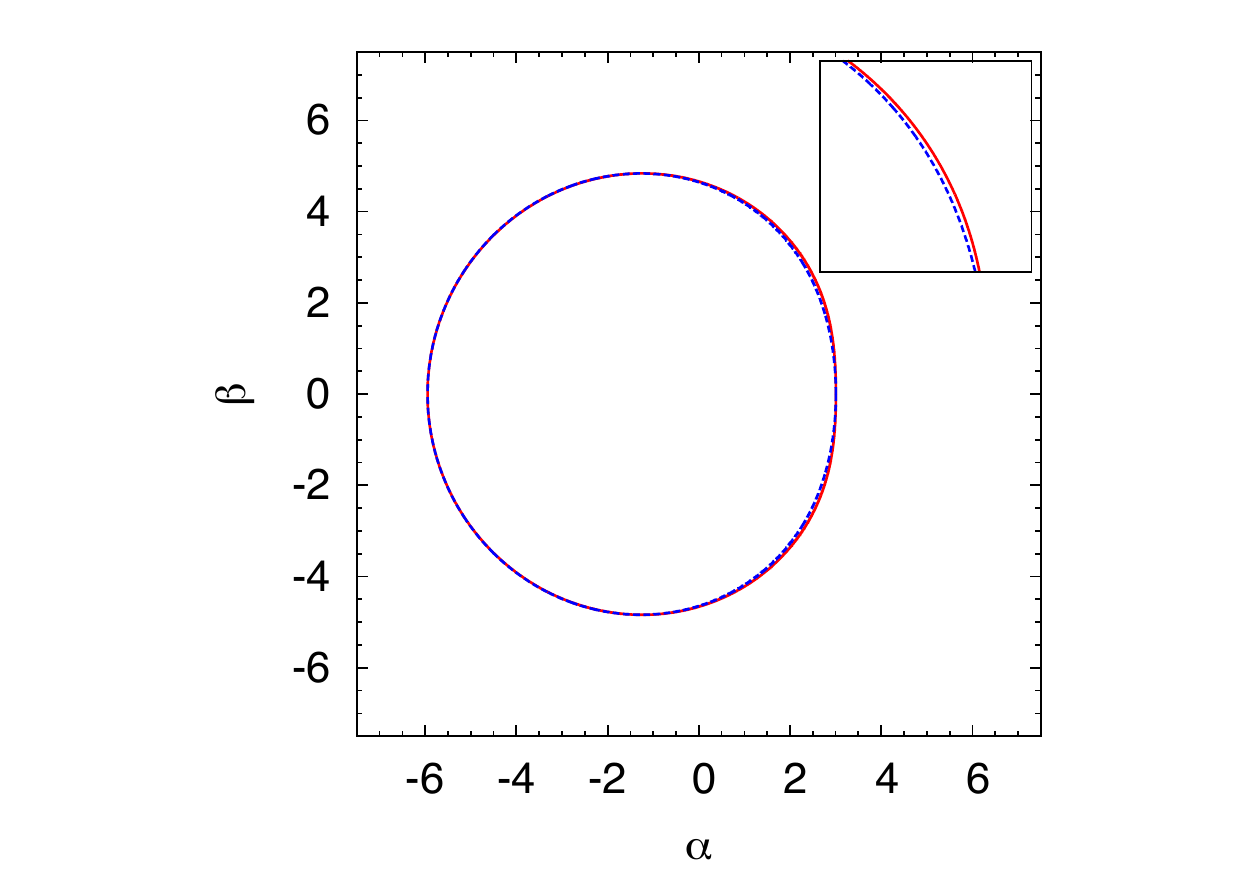}
\end{center}
\vspace{-0.5cm}
\caption{Left panel: shadows of a Kerr BH (blue dashed line) and of a Bardeen 
BH with $g/M = 0.6$ (red solid line) with the same mass $M = 1$, the same 
viewing angle $i = 90^\circ$, and the same Hioki-Maeda distortion parameter 
$\delta = 0.1500$. The Kerr BH has $a/M = 0.9189$, while the Bardeen BH has
$a/M = 0.5286$. Right panel: as in the left panel, but for two BHs with the same
shadow radius $R$ on the sky ($M = 0.9311$ for the Kerr BH, $M=1$ for the 
Bardeen one) and the same position of the center of the circle $C$ (the shadow
of the Kerr BH has been shifted by 0.433 along the $\alpha$ direction). See the 
text for more details.}
\label{fig3}
\end{figure}

\section{Measuring the spin and the deformation parameters from the observation of the shadow \label{s-def}}

Since the Hioki-Maeda distortion parameter $\delta$ is degenerate with respect 
to the BH spin and possible deviations from the Kerr solution (and it could not 
be otherwise, because one parameter of the shadow can at most be used to 
determine one parameter of the background geometry), in this section we look for the 
best choice of the second parameter to test the Kerr metric. If we consider non-rotating
BHs, the shadow is a circle for any value of $g/M$ and the sole difference is its
radius, which depends on $g/M$. 
For $a/M \neq 0$, we can expect that shadows with the same Hioki-Maeda
distortion parameter have different shape and that the difference increases
as $g/M$ and $a/M$ increase and $i$ approaches $90^\circ$. As first step, it is 
useful to visualize such a difference. This is done in Fig.~\ref{fig3}, where we 
compare the shadows of a Kerr BH and of a Bardeen BH with $g/M = 0.6$ for 
$i = 90^\circ$. We start from imposing that $\delta = 0.1500$. We find that such
a distortion parameter corresponds to a Kerr BH with $a/M = 0.9189$ and to a
$g/M = 0.6$ Bardeen BH with $a/M = 0.5286$. The left panel shows the two 
shadows as computed using the same BH mass $M$. The right panel 
compares the two shadows after properly rescaling the one of the Kerr BH
and shifting it on the celestial plane, so that their radii have the same value
and the centers of the shadows coincide. We note that the maximum value
of the spin for a Bardeen BH with $g/M = 0.6$ is $a/M \approx 0.5295$, so 
we are considering an almost extremal BH.

\begin{figure}
\begin{center}
\hspace{-1cm}
\includegraphics[type=pdf,ext=.pdf,read=.pdf,width=8.7cm]{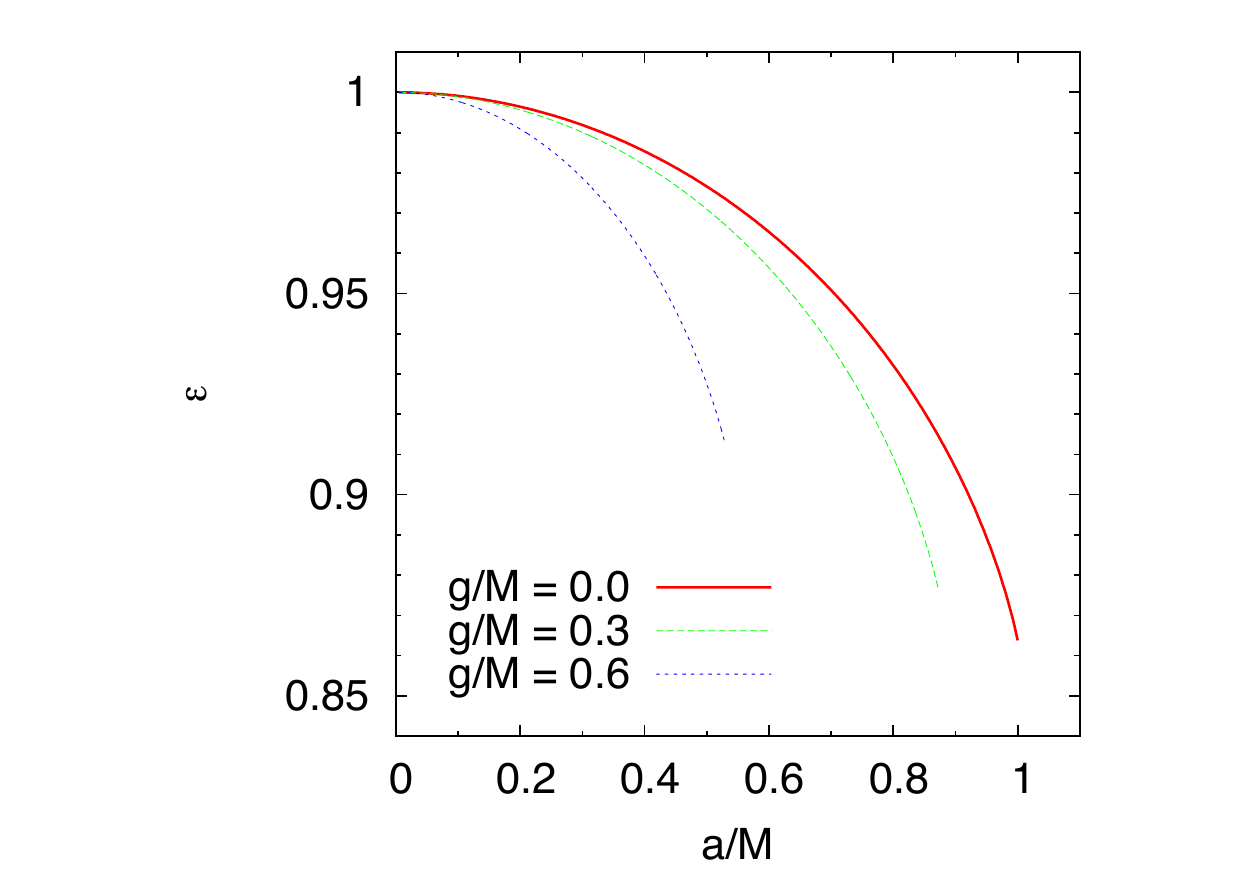}
\hspace{-1.5cm}
\includegraphics[type=pdf,ext=.pdf,read=.pdf,width=8.7cm]{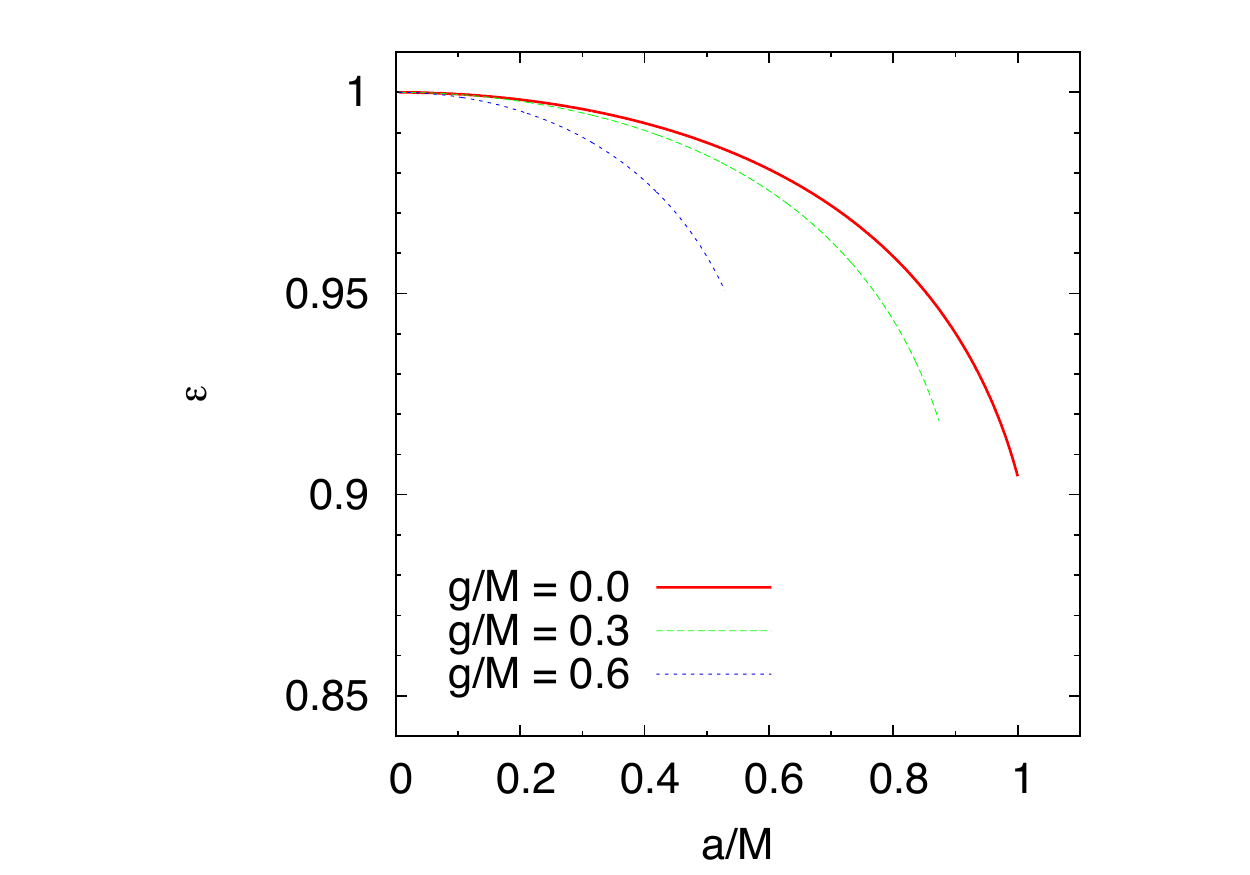}
\end{center}
\vspace{-0.5cm}
\caption{Distortion parameter $\epsilon$ as a function of the spin
parameter $a/M$ for Kerr BHs (red solid line), Bardeen BHs with $g/M=0.3$
(green dashed line), and Bardeen BHs with $g/M=0.6$ (blue dotted line).
The inclination angle is $i = 90^\circ$ (left panel) and $45^\circ$ (right panel).}
\label{fig4}
\end{figure}

\subsection{Distortion parameter $\epsilon$}

From the right panel in Fig.~\ref{fig3}, we can realize that the main difference
in the shadow shape is in the apparent photon capture radii on the right
side, corresponding to the ones associated to corotating orbits. We are thus tempted
to defined the distortion parameter $\epsilon$ as follows. With reference to
Fig.~\ref{fig1}, we call $S$ the distance between the center of the circle of 
the shadow, $C$, and the point on the right side of the boundary
with coordinate $\beta = \beta_{\rm max}/2$, where $\beta_{\rm max}$ is the 
$\beta$ coordinate of the top end of the shadow used to find $R$. We then
define $\epsilon = S/R$ which, like the Hioki-Maeda distortion parameter
$\delta$, only depends on the shape of the shadow. The distortion parameter 
$\epsilon$ as a function of the spin parameter $a/M$ for Kerr BHs, 
Bardeen BHs with $g/M=0.3$, and Bardeen BHs with $g/M=0.6$ is shown 
in Fig.~\ref{fig4}.

\begin{figure}
\begin{center}
\hspace{-1cm}
\includegraphics[type=pdf,ext=.pdf,read=.pdf,width=8.7cm]{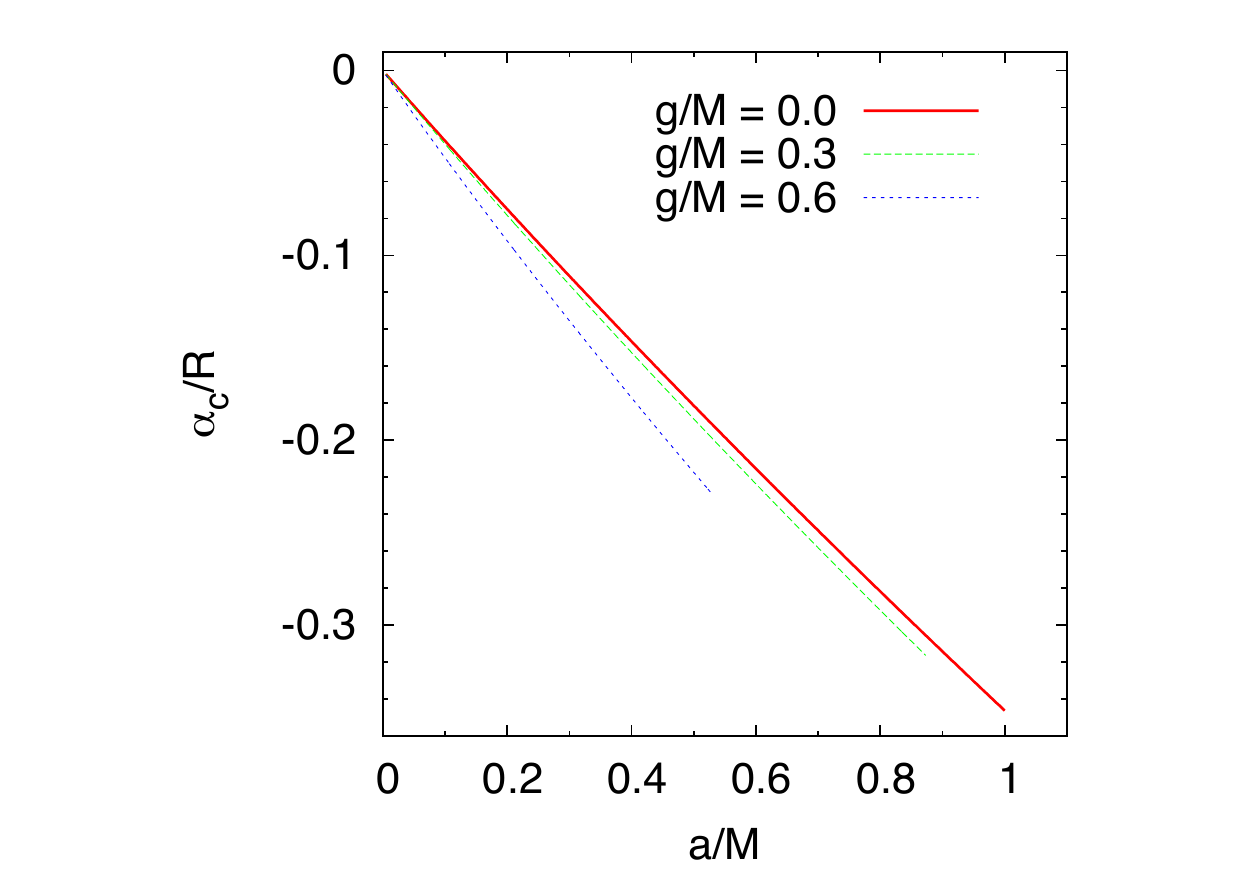}
\hspace{-1.5cm}
\includegraphics[type=pdf,ext=.pdf,read=.pdf,width=8.7cm]{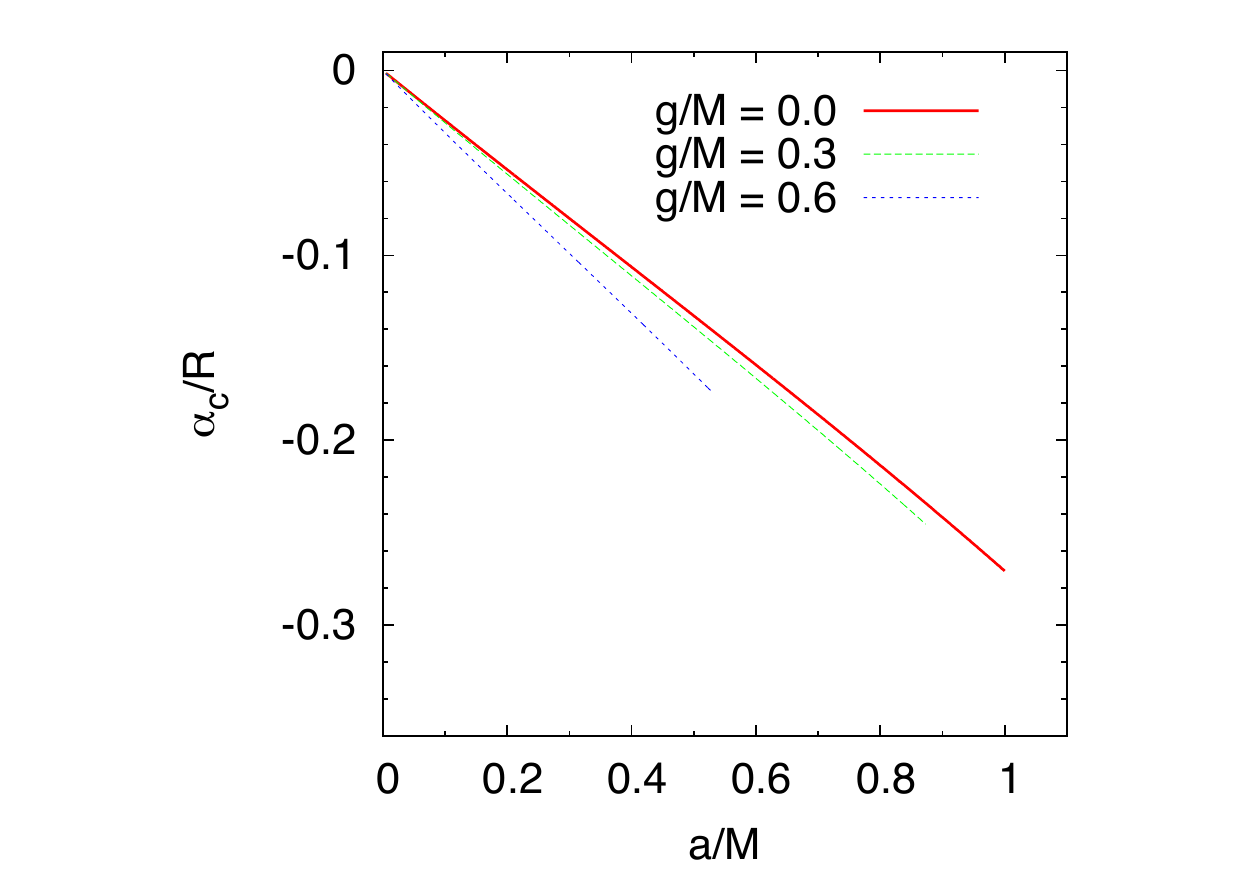}
\end{center}
\vspace{-0.5cm}
\caption{Position of the center of the circle on the sky as a function of the spin
parameter $a/M$ for Kerr BHs (red solid line), Bardeen BHs with $g/M=0.3$
(green dashed line), and Bardeen BHs with $g/M=0.6$ (blue dotted line).
The inclination angle is $i = 90^\circ$ (left panel) and $45^\circ$ (right panel).}
\label{fig5}
\end{figure}

\subsection{Position of the center of the shadow}

The shapes of the shadows of Kerr and Bardeen BHs are clearly very similar 
and therefore only very accurate image can distinguish the two metrics and 
provide a meaningful constraint on $g/M$. One can therefore try to follow a 
different strategy from the exact determination of the shadow shape. The left 
panel in Fig.~\ref{fig3} may suggest that this could be achieved from the estimate 
of the exact position on the sky of the center of the circle, $C$, with respect to the
actual center of the system, $\alpha = \beta = 0$. In principle, that is surely an 
available option and Fig.~\ref{fig5} shows $\alpha_{\rm c}/R$ as a function
of the spin parameter for the shadows of Kerr and Bardeen BHs. We have
plotted $\alpha_{\rm c}/R$ instead of $\alpha_{\rm c}$ or $\alpha_{\rm c}/M$ 
because in this way we do not assume an accurate measurement of the BH 
mass and distance. We just need a very good measurement of the BH position 
on the sky.

\begin{figure}
\begin{center}
\hspace{-1cm}
\includegraphics[type=pdf,ext=.pdf,read=.pdf,width=8.7cm]{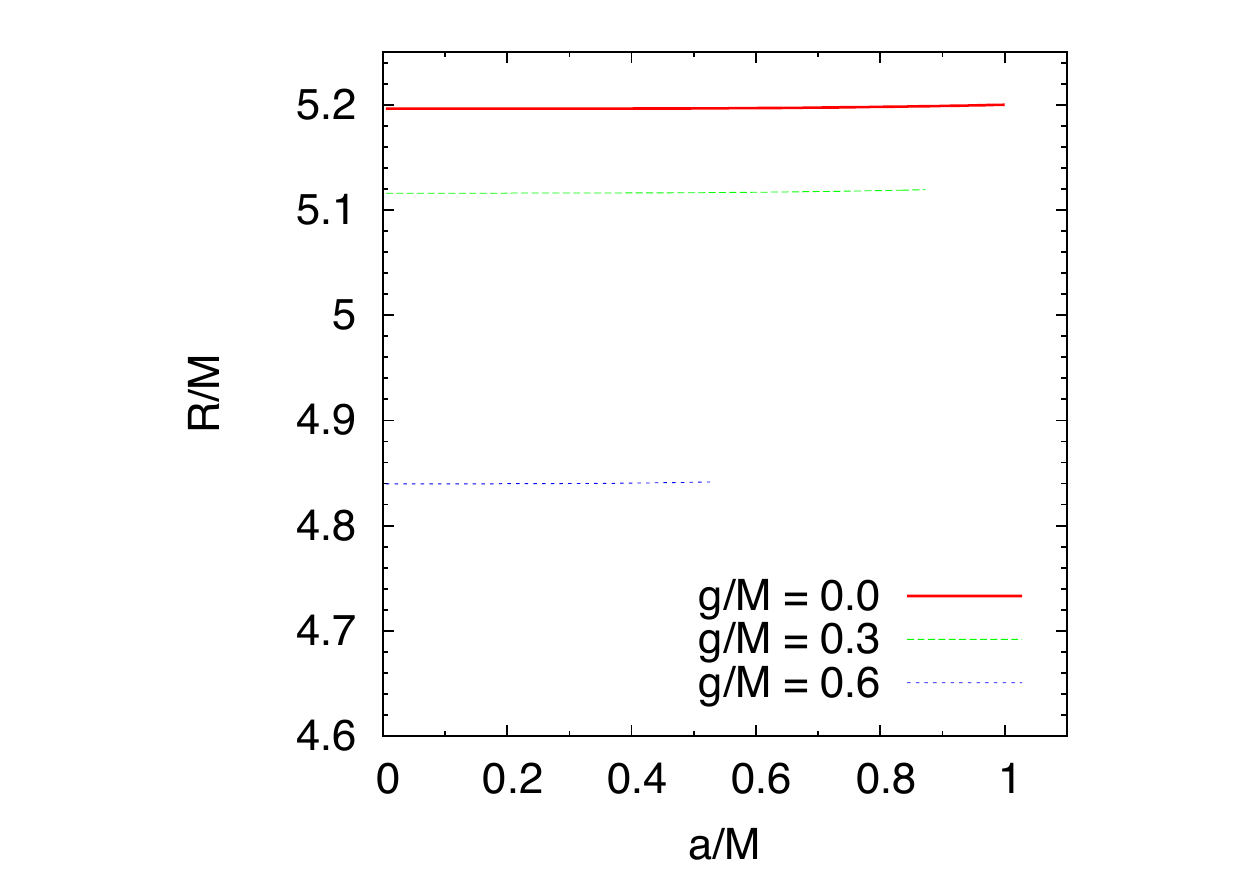}
\hspace{-1.5cm}
\includegraphics[type=pdf,ext=.pdf,read=.pdf,width=8.7cm]{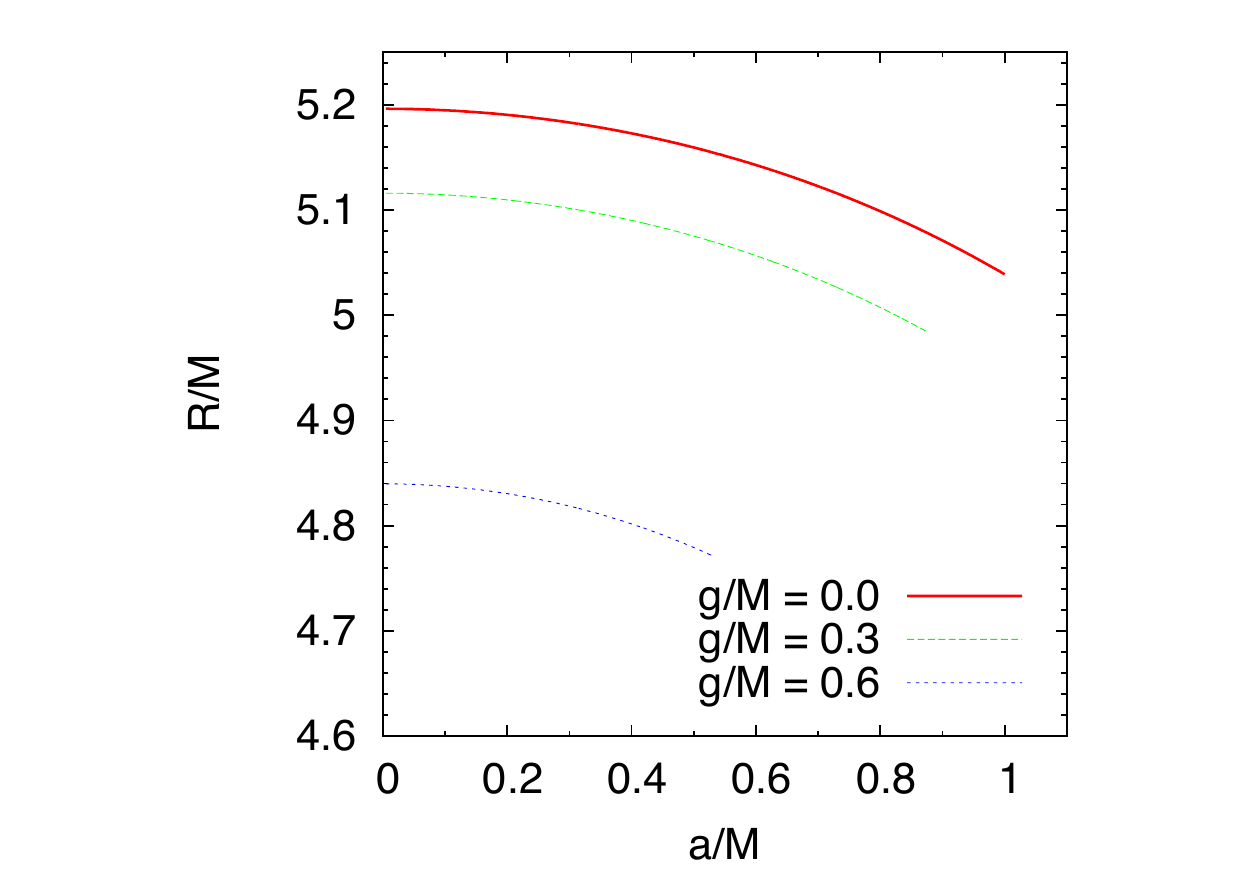}
\end{center}
\vspace{-0.5cm}
\caption{Radius of the shadow $R/M$ as a function of the spin
parameter $a/M$ for Kerr BHs (red solid line), Bardeen BHs with $g/M=0.3$
(green dashed line), and Bardeen BHs with $a/M=0.6$ (blue dotted line).
The inclination angle is $i = 90^\circ$ (left panel) and $45^\circ$ (right panel).}
\label{fig6}
\end{figure}

\subsection{Radius of the shadow $R$}

Lastly, we consider the possibility that we can get very good estimates of the 
BH mass and distance and we can therefore combine the Hioki-Maeda 
distortion parameter $\delta$ with the measurement of the radius of the 
shadow $R$. $R/M$ as a function of the spin parameter $a/M$ for 
$g/M = 0.0$ (Kerr), 0.3, and 0.6 is shown in Fig.~\ref{fig6}. It is remarkable
that for an observer near the equatorial plane the value of $R/M$ is mainly
determined by $g/M$ and it is not very sensitive to the spin. The dependence
of $R$ on the spin increases as the observer moves towards the axis of 
symmetry, but it is still weak for $i = 45^\circ$, as shown the right 
panel in Fig.~\ref{fig6}.

\begin{figure}
\begin{center}
\hspace{-1cm}
\includegraphics[type=pdf,ext=.pdf,read=.pdf,width=8.7cm]{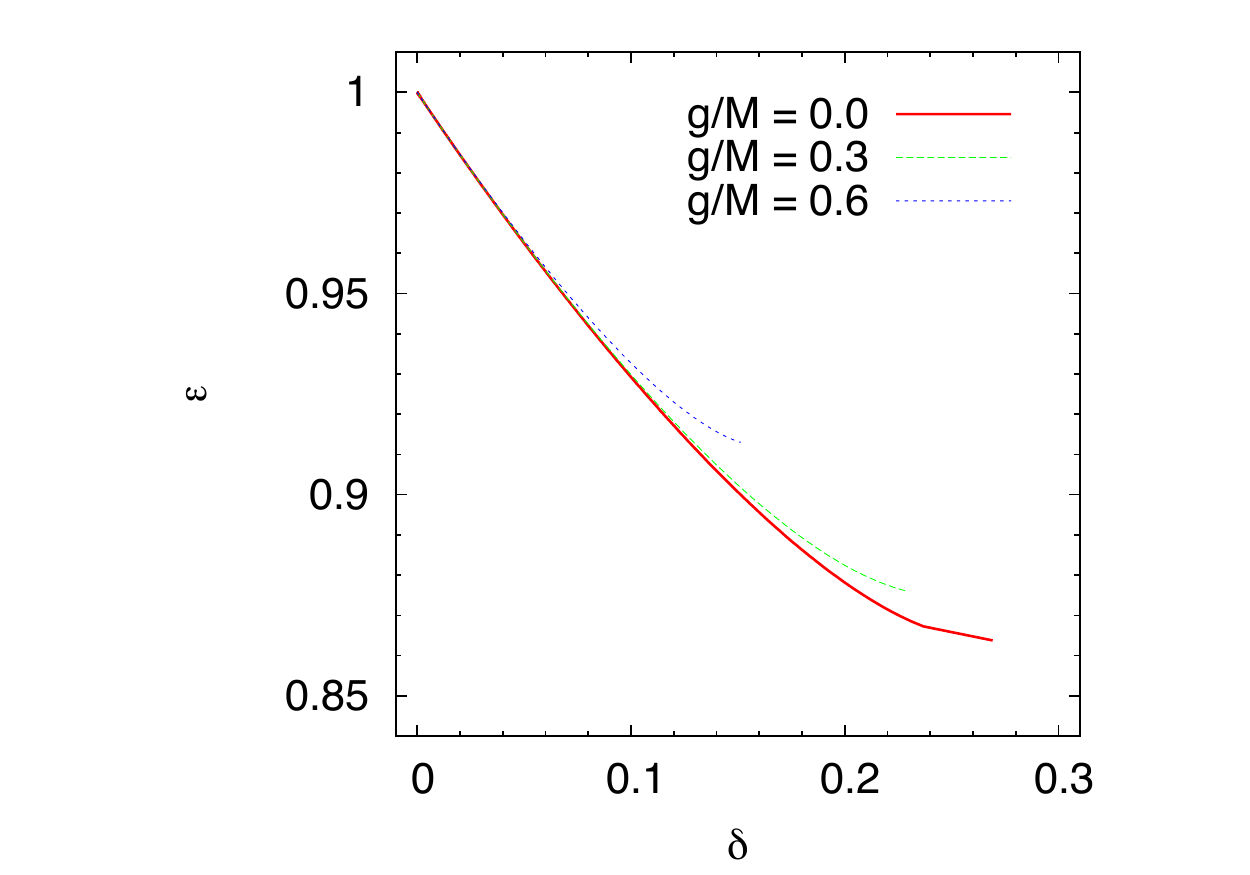}
\hspace{-1.5cm}
\includegraphics[type=pdf,ext=.pdf,read=.pdf,width=8.7cm]{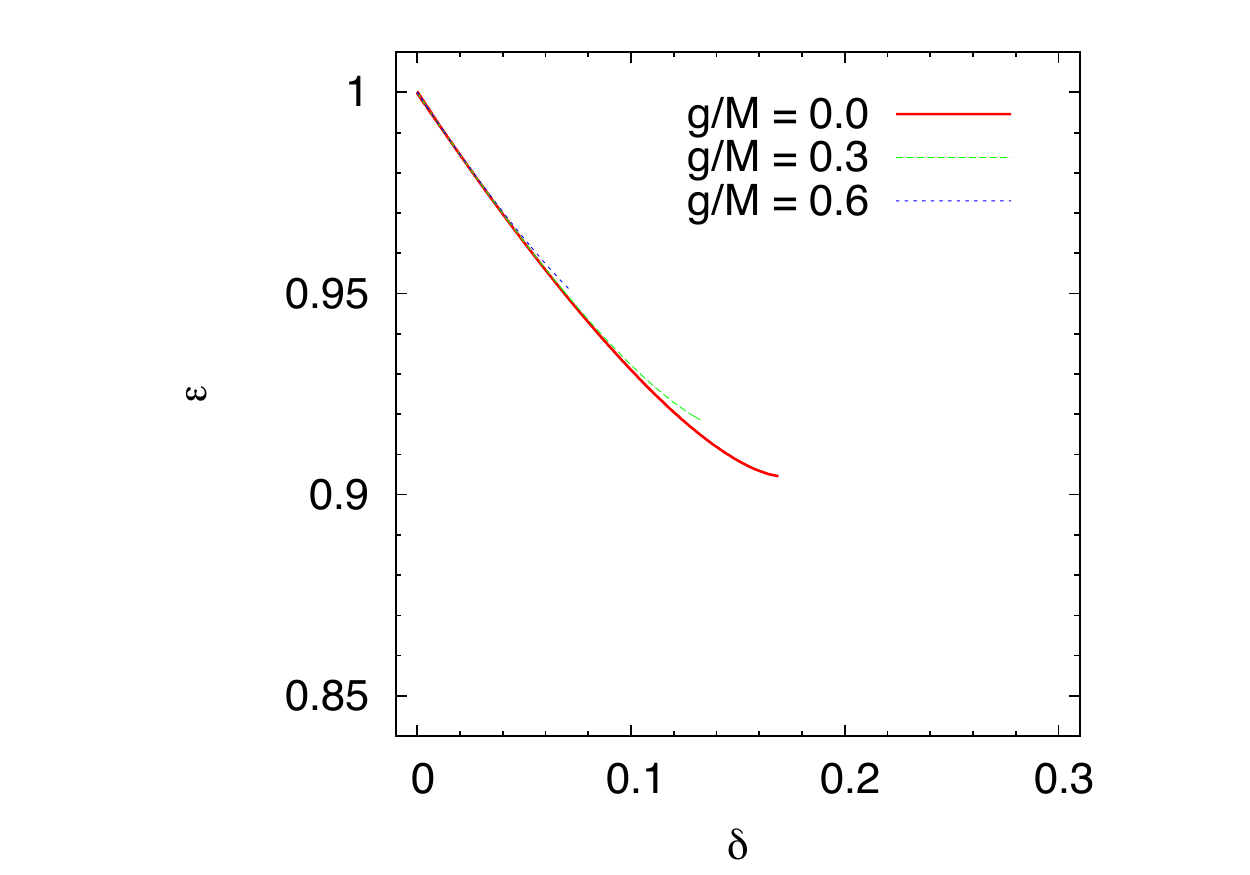}
\end{center}
\vspace{-0.5cm}
\caption{The Hioki-Maeda distortion parameter $\delta$ against the distortion
parameter $\epsilon$ for Kerr BHs (red solid line), Bardeen BHs with $g/M=0.3$
(green dashed line), and Bardeen BHs with $a/M=0.6$ (blue dotted line).
The inclination angle is $i = 90^\circ$ (left panel) and $45^\circ$ (right panel).}
\label{fig7}
\end{figure}

\begin{figure}
\begin{center}
\hspace{-1cm}
\includegraphics[type=pdf,ext=.pdf,read=.pdf,width=8.7cm]{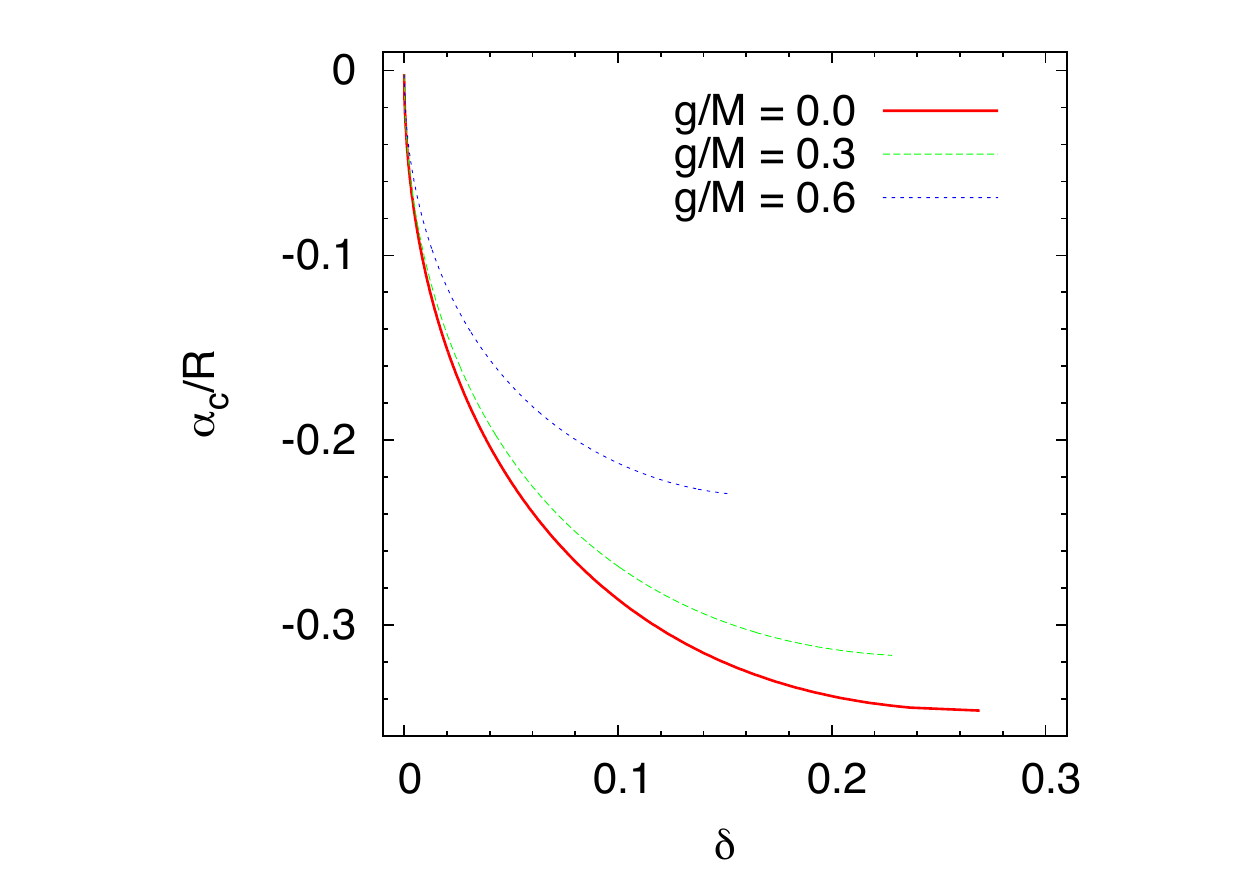}
\hspace{-1.5cm}
\includegraphics[type=pdf,ext=.pdf,read=.pdf,width=8.7cm]{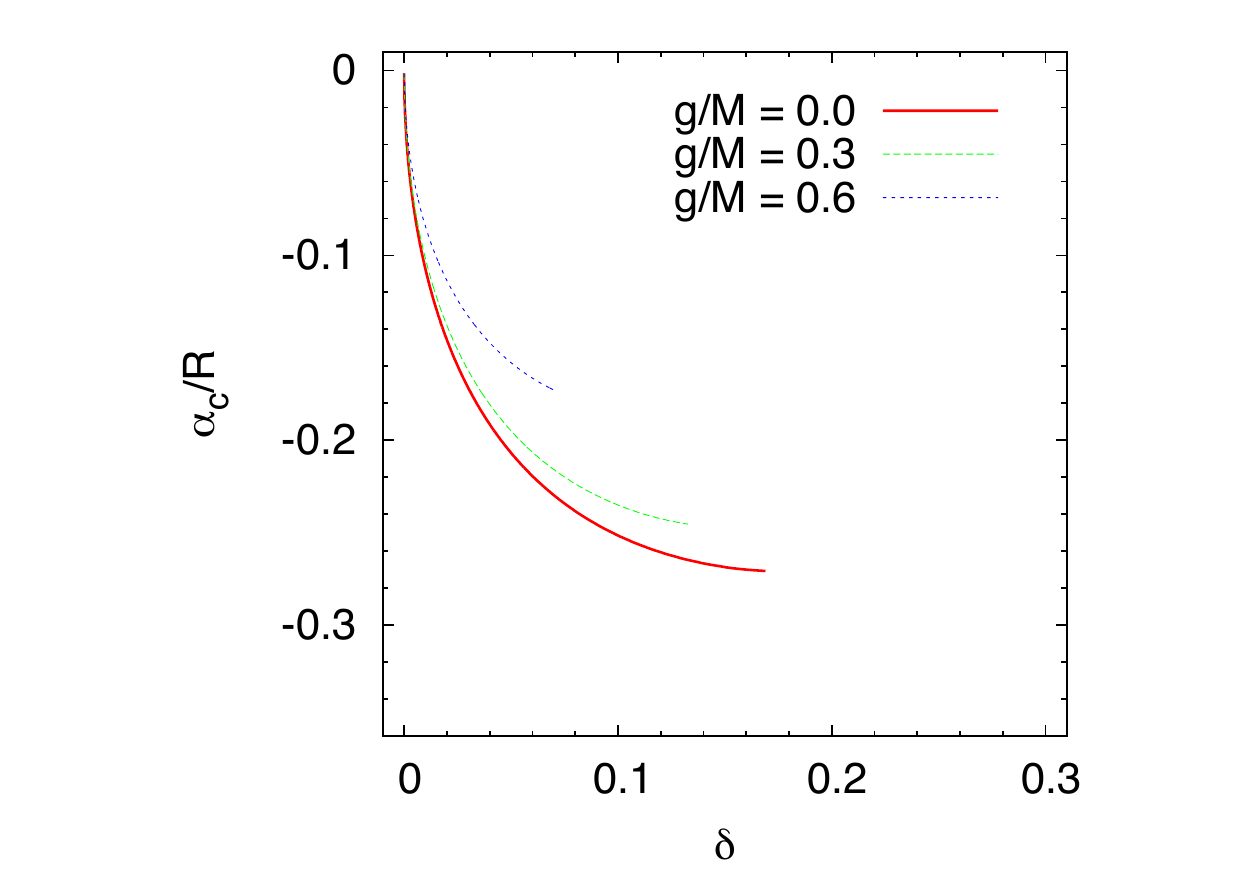}
\end{center}
\vspace{-0.5cm}
\caption{The Hioki-Maeda distortion parameter $\delta$ against the position of 
the center of the circle on the sky for Kerr BHs (red solid line), Bardeen BHs with 
$g/M=0.3$ (green dashed line), and Bardeen BHs with $a/M=0.6$ (blue dotted line).
The inclination angle is $i = 90^\circ$ (left panel) and $45^\circ$ (right panel).}
\label{fig8}
\end{figure}

\begin{figure}
\begin{center}
\hspace{-1cm}
\includegraphics[type=pdf,ext=.pdf,read=.pdf,width=8.7cm]{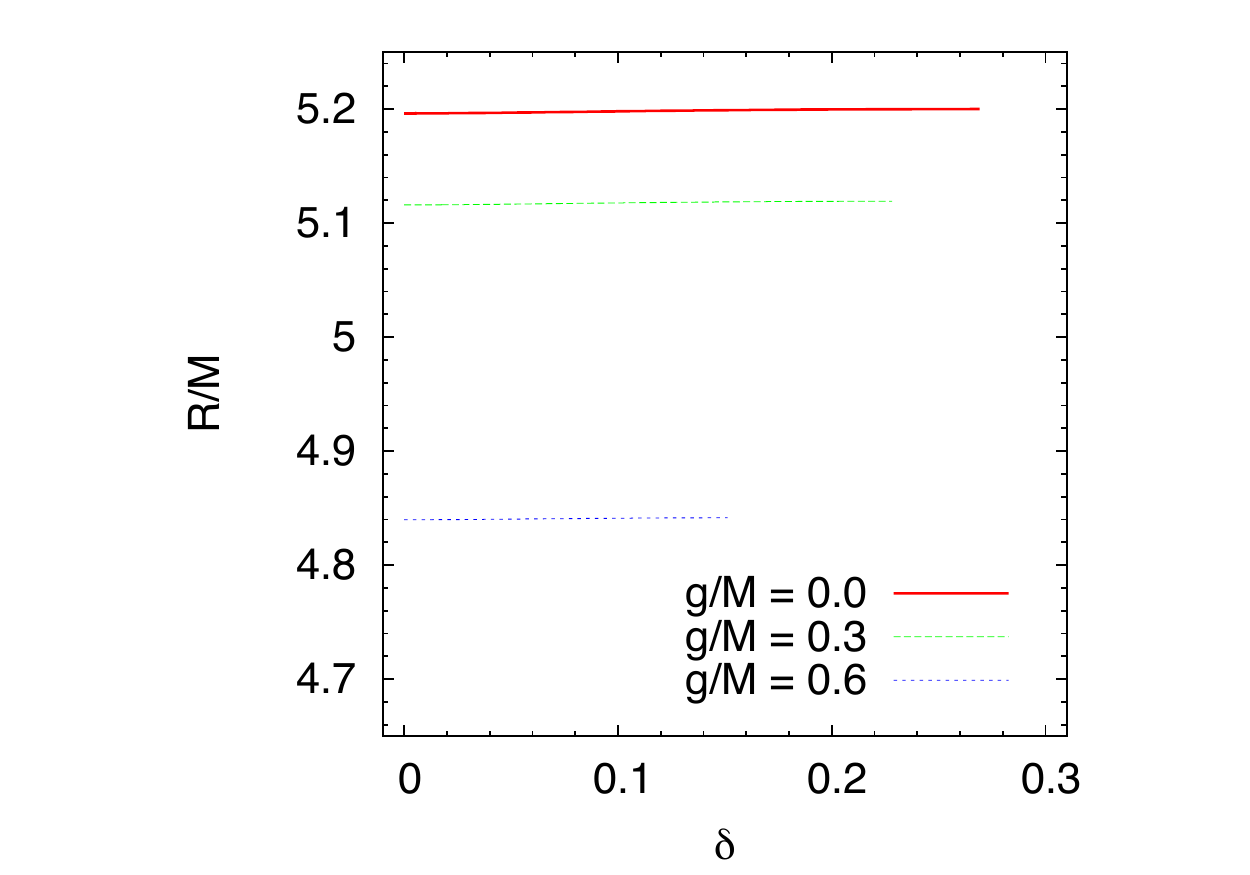}
\hspace{-1.5cm}
\includegraphics[type=pdf,ext=.pdf,read=.pdf,width=8.7cm]{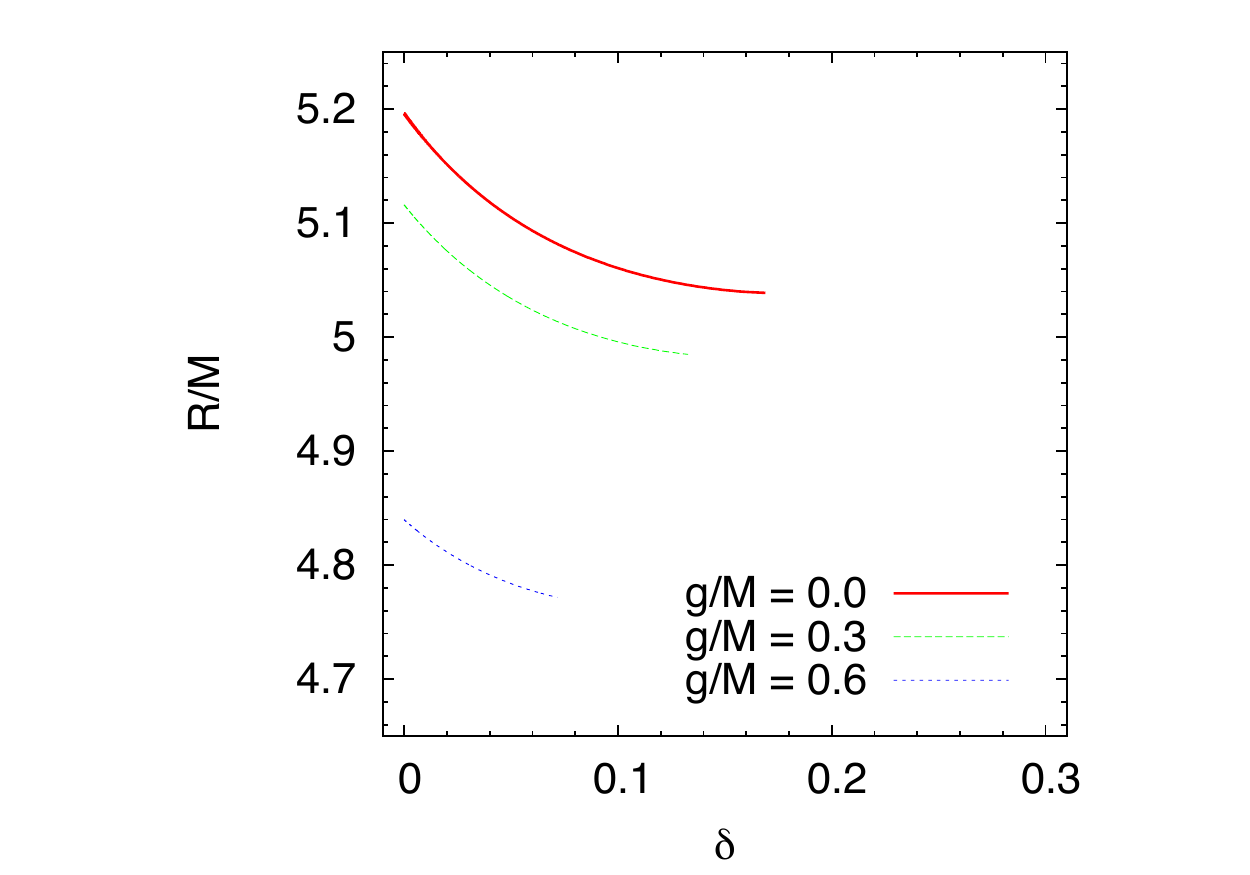}
\end{center}
\vspace{-0.5cm}
\caption{The Hioki-Maeda distortion parameter $\delta$ against the radius of 
the shadow $R/M$ for Kerr BHs (red solid line), Bardeen BHs with $g/M=0.3$ 
(green dashed line), and Bardeen BHs with $a/M=0.6$ (blue dotted line).
The inclination angle is $i = 90^\circ$ (left panel) and $45^\circ$ (right panel).}
\label{fig9}
\end{figure}

\begin{figure}
\begin{center}
\hspace{-1cm}
\includegraphics[type=pdf,ext=.pdf,read=.pdf,width=8.7cm]{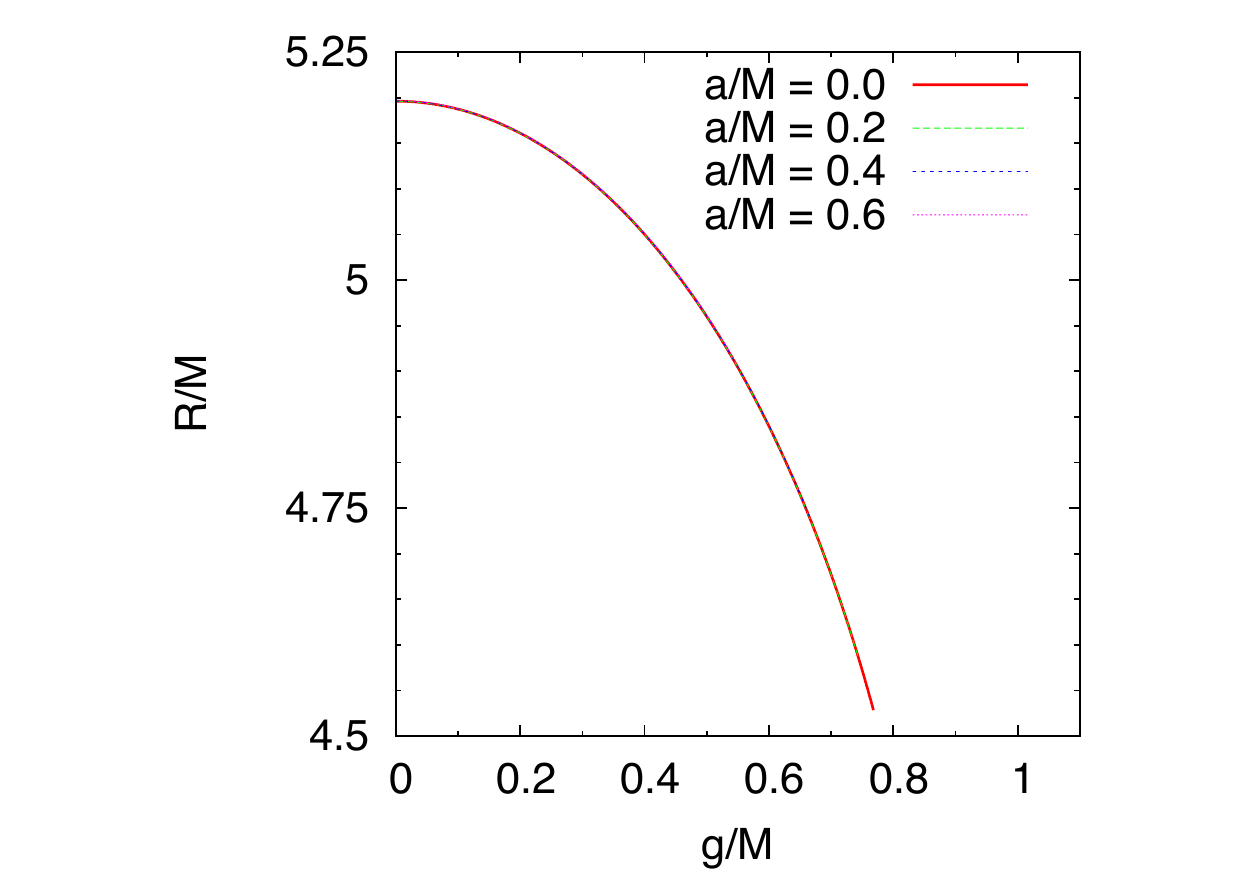}
\hspace{-1.5cm}
\includegraphics[type=pdf,ext=.pdf,read=.pdf,width=8.7cm]{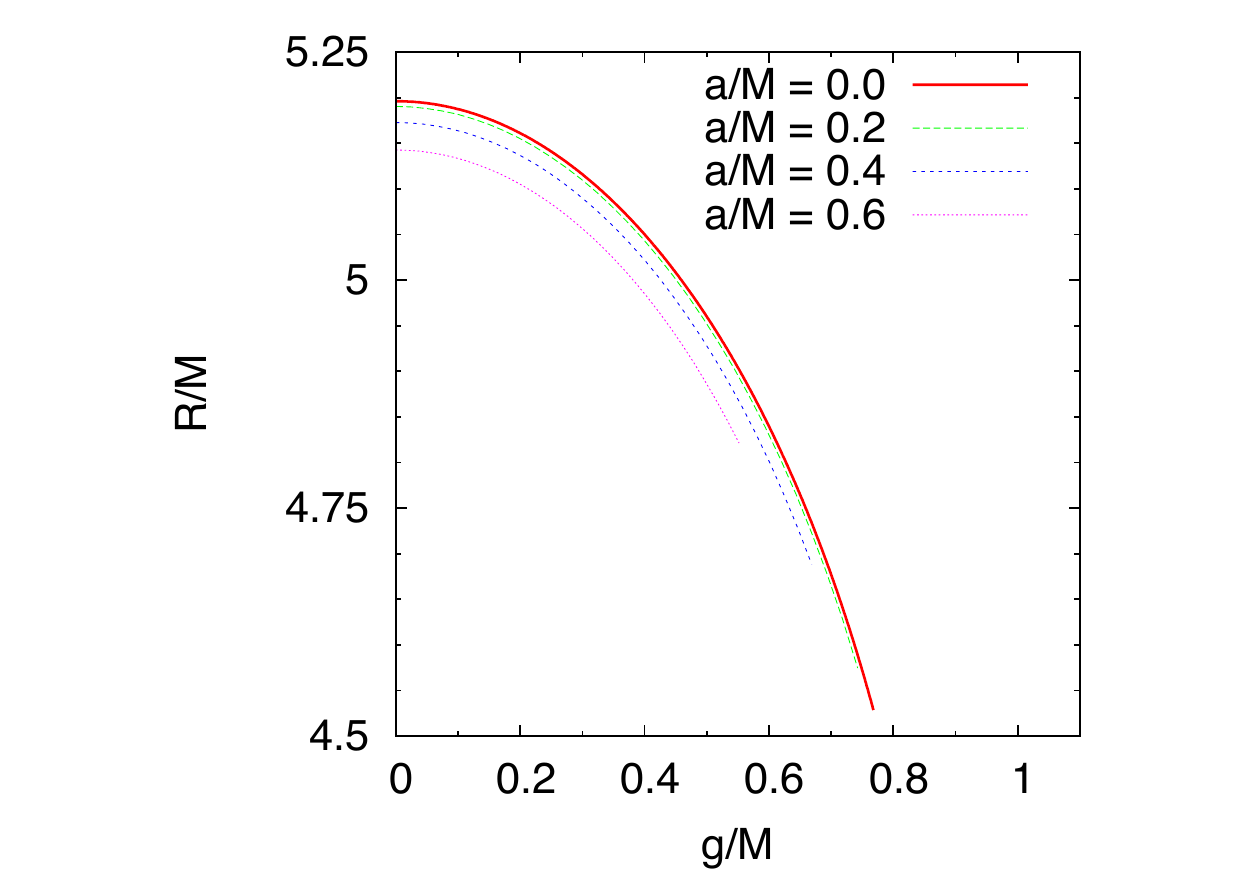}\\
\includegraphics[type=pdf,ext=.pdf,read=.pdf,width=8.7cm]{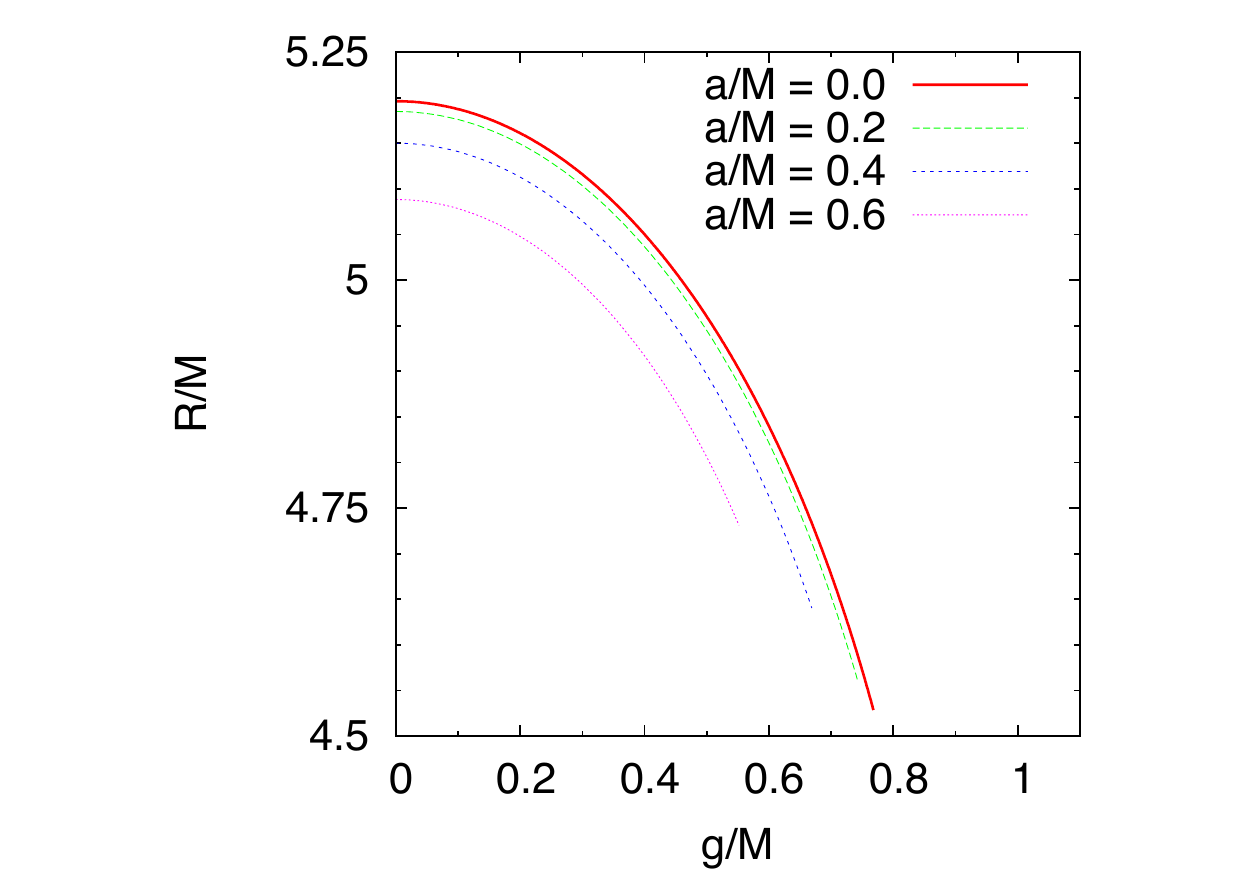}
\end{center}
\vspace{-0.5cm}
\caption{The radius of the shadow $R/M$ as a function of the Bardeen charge 
$g/M$ for different values of the spin parameter and an observer inclination
angle $i = 90^\circ$ (top left panel), $45^\circ$ (top right panel), and 
$10^\circ$ (bottom panel).}
\label{fig10}
\end{figure}

\section{Discussion \label{s-d}}

At least in principle, the simultaneous measurement of the Hioki-Maeda distortion
parameter $\delta$ and one of the three parameters discussed in the previous 
section ($\epsilon$, $\alpha_{\rm c}$, or $R$) breaks the degeneracy between
the BH spin and possible deviations from the Kerr geometry and it is therefore 
a potential approach to test the Kerr metric around SgrA$^*$ from the observation
of its shadow. This is more evident if we plot the Hioki-Maeda distortion parameter 
against one of the other three. The plots are shown in Fig.~\ref{fig7} for the distortion 
parameter $\epsilon$, in Fig.~\ref{fig8} for the shadow center $\alpha_{\rm c}/M$, 
and in Fig.~\ref{fig9} for the shadow radius $R/M$. The fact that these curves 
depend on the value of $g/M$ is enough to conclude that there is no degeneracy.
However, the above consideration is correct only in principle, in the sense that 
we have also to check if it is possible to measure $\epsilon$, $\alpha_{\rm c}$, 
or $R$ with sufficient precision to distinguish Kerr and Bardeen BHs.

In the case of the distortion parameter $\epsilon$, the difference between Kerr
and Bardeen BHs is clearly tiny, as it was already evident from Fig.~\ref{fig3}.
Even in the most optimistic case of an inclination angle $i = 90^\circ$, for the
same Hioki-Maeda distortion parameter $\delta$ the difference between the
parameter $\epsilon$ for Kerr BHs and Bardeen BHs with $g/M = 0.3$ is never 
larger than 0.8\%, lower 0.2\% for $\delta < 0.15$, and lower than 0.08\% for 
$\delta < 0.10$. If we compare Kerr BHs and Bardeen BHs with $g/M = 0.6$, 
we find that shadows with the same $\delta$ have $\epsilon$ parameters that 
differ less than 1.5\% (but it is close to 1.5\% only when the Bardeen BH is an
almost extremal object). Such precisions are at least extremely challenging, 
even in the most favorable conditions for $a/M$, $g/M$, and $i$. The 
uncertainties on $\delta$ and $\epsilon$ are
\be
\frac{\Delta \delta}{\delta} = \frac{\Delta R}{R} + \frac{\Delta D}{D} \, , \quad
\frac{\Delta \epsilon}{\epsilon} = \frac{\Delta R}{R} + \frac{\Delta S}{S} \, ,
\ee
where $\Delta R$, $\Delta D$, and $\Delta S$ are the uncertainties on $R$, 
$D$, and $S$. For SgrA$^*$, we expect $R \approx 30$~$\mu$as. With an
imaging resolution of $\sim 0.3$~$\mu$as, $\Delta \epsilon/\epsilon$ is
already 2\% or worst, which is not enough even to distinguish a Kerr from a
Bardeen BH with $g/M = 0.6$. Here we are also assuming to know the
inclination angle $i$ with arbitrary precision, which is surely not the case
and its uncertainty introduces an additional error in the estimate of 
$\epsilon$. The fact that the shapes of the shadows in Kerr 
and non-Kerr spacetimes are usually very similar, and therefore difficult to
distinguish with observations, was already stressed in Ref.~\cite{sh5} from 
the comparison of Kerr and Tomimatsu-Sato shadows. Other non-Kerr 
metrics might have shadows with more significant deviations, but more often
the difference seems to be very small.

The measurement of the shift between the position of the center of the shadow 
and the actual center of the system may initially look more promising, because 
Fig.~\ref{fig8} shows that the lines for different values of $g/M$ have a large 
separation. Unfortunately, the position of the center of the system on the
sky is very difficult to determine and the uncertainty is $\Delta \alpha_{\rm c} \sim 1$~mas
(but see Ref.~\cite{position}, where a position relative to a reference point with
an accuracy of order 1~$\mu$as might be possible).

The last parameter of the shadow discussed in the previous section is the 
apparent size of the radius $R$ (see Figures~\ref{fig9} and \ref{fig10}). 
In this case, we need very good measurement 
of the BH mass and distance from us. At present, these quantities are difficult 
to measure and the final uncertainty on the expect apparent size on the sky
of $R$ is around 15\%~\cite{sh11}. Such an uncertainty is larger than the 
difference between the radius of the shadow of a Kerr BH and of a Bardeen 
BH with $g/M = 0.6$ with the same $\delta$, which is around 7\% for 
$i = 90^\circ$ and independently of the spin parameter $a/M$.

The measurement of the radius of the shadow may turn out to be the most
promising approach in the case of significant improvements of the estimate
of the BH mass and of our distance from the galactic center. That could be
achieved in the case of the discovery of a radio pulsar in a compact orbit
(i.e. with an orbital periods of a few months) around SgrA$^*$. As discussed 
in Ref.~\cite{pulsar}, pulsar timing can determine the Keplerian and Post-Keplerian
parameters and therefore get a robust estimate of the BH mass, independently
of the distance from us from the galactic center, which can instead be inferred
with high precision by combining the BH mass with near-infrared astrometric 
measurements. In this case, the uncertainty on $R$ would be determined
by the imaging resolution. If the resolution is at the level of 0.3~$\mu$as, 
$R$ can be measured with a precision of 1\%. If SgrA$^*$ is rotating rapidly,
after a few years of observations of the radio pulsar, the spin parameter 
$a/M$ could be determined with a precision of order 0.1\% (but the 
uncertainty would be significantly larger for a mid-rotating or slow-rotating 
BH)~\cite{pulsar}. Such a measurement could also be combined with the ones of $\delta$
and $R$ on the spin-Bardeen charge plane. Since the pulsar is relatively
far from the BH, the pulsar measurement of the frame dragging is really sensitive 
to the value of $a/M$, independently of the nature if the SgrA$^*$, because possible 
deviations from the Kerr solution are suppressed by powers in $M/r \ll 1$, where 
$r$ is the distance of the pulsar\footnote{While pulsar timing can also 
measure the BH quadrupole moment and thus test the Kerr nature of 
SgrA$^*$ (at least in the case of a fast-rotating BH)~\cite{pulsar}, the 
combination of the measurement of the shadow and of the pulsar can test 
if there are deviations of higher order.}.

\begin{figure}
\begin{center}
\hspace{-1cm}
\includegraphics[type=pdf,ext=.pdf,read=.pdf,width=8.7cm]{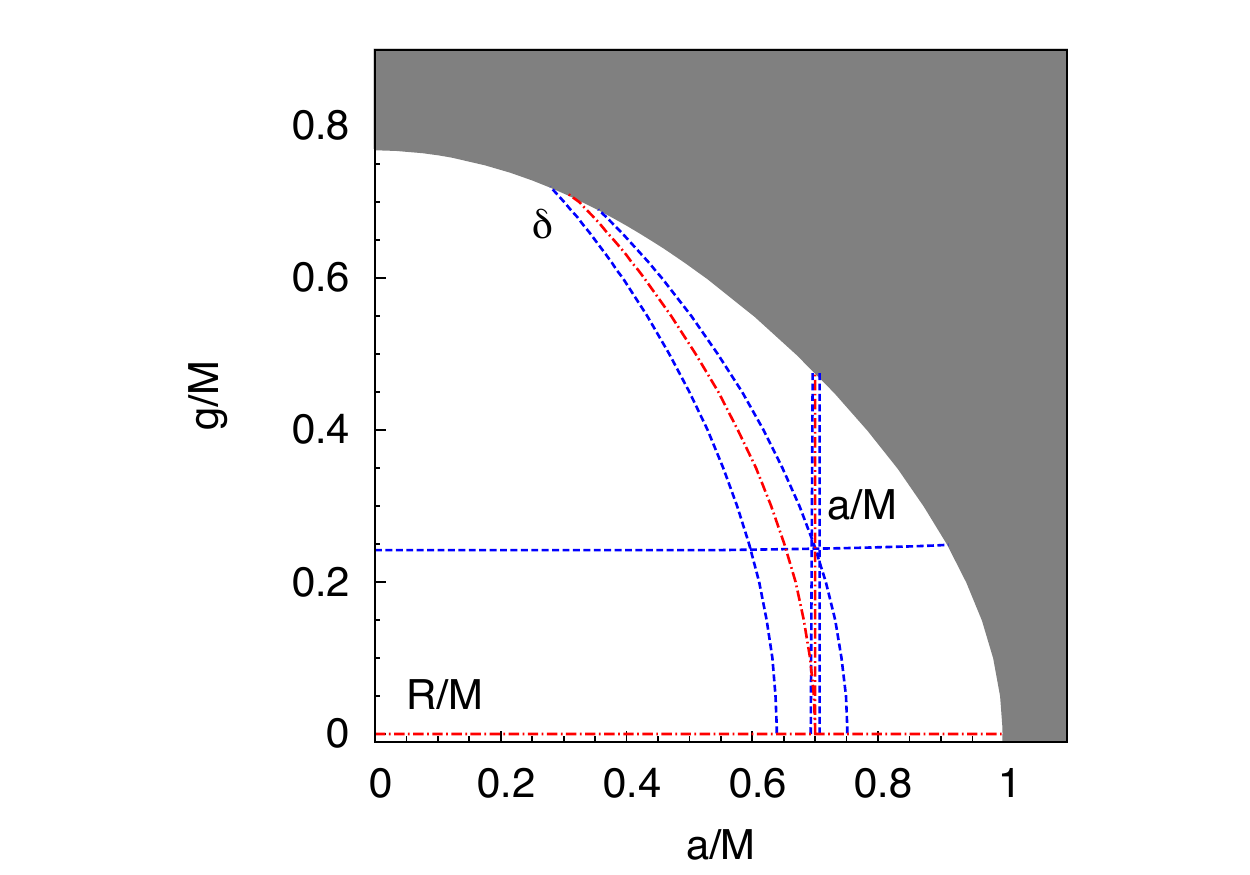}
\hspace{-1.5cm}
\includegraphics[type=pdf,ext=.pdf,read=.pdf,width=8.7cm]{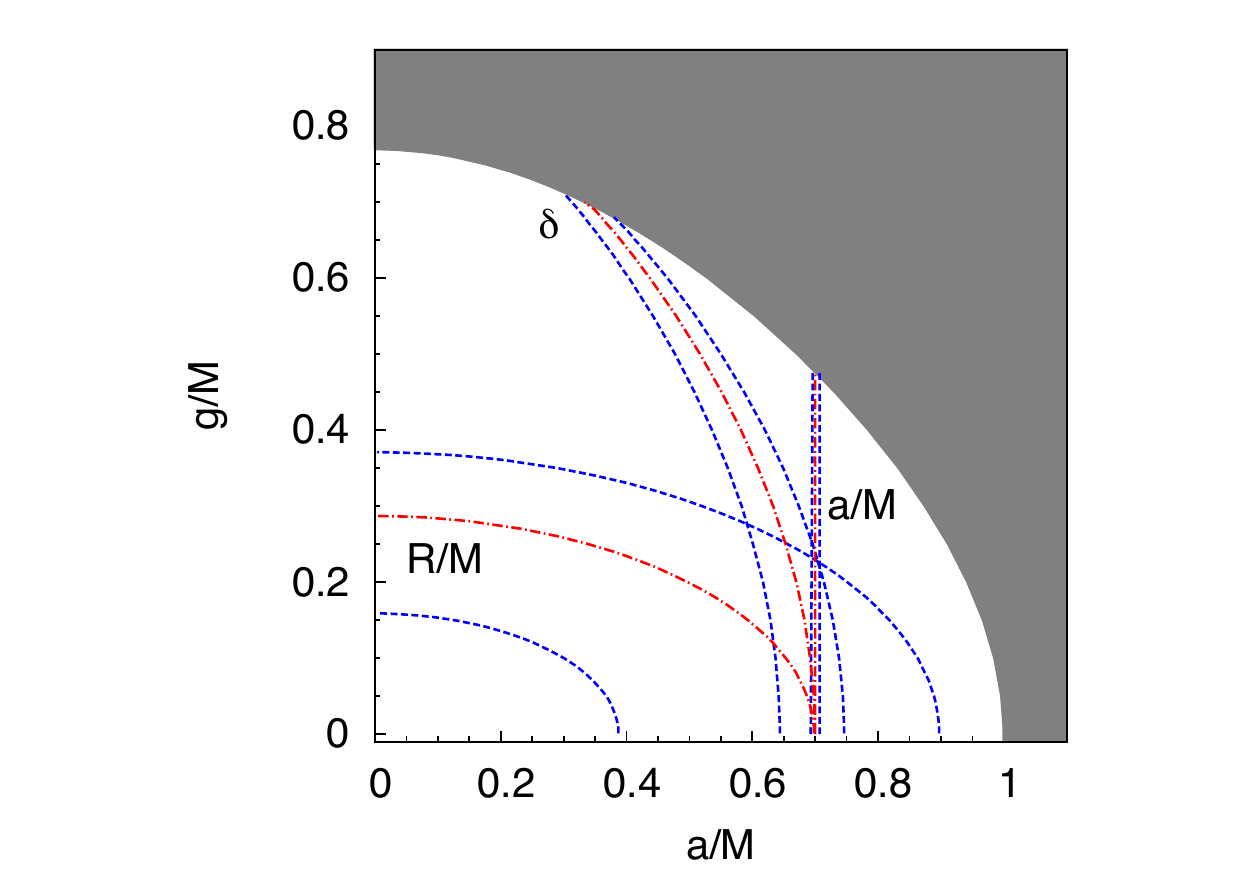}
\end{center}
\vspace{-0.5cm}
\caption{Hypothetical constraints from the measurements of the Hioki-Maeda 
distortion parameter $\delta$ determined with a precision of 20\%, of the shadow
radius $R/M$ determined with a precision of 1\%, and of the spin parameter $a/M$
inferred from the orbital motion of a pulsar in a compact orbit and determined 
with a precision of 1\%, assuming that the object is a Kerr BH with $a/M = 0.7$
and the inclination angle is $i = 90^\circ$ (left panel, in this case $\delta = 0.0668$
and $R/M = 5.20$) and $45^\circ$ (right panel, in this case $\delta = 0.0371$
and $R/M = 5.12$). The red dashed-dotted lines are the central values of the 
measurements, while the blue dashed curves correspond to their uncertainties.
The gray area is the region of objects without event horizon and can be ignored.}
\label{fig11}
\end{figure}

The possible constraints from the simultaneous measurements of the 
Hioki-Maeda distortion parameter $\delta$, the shadow radius $R/M$, and the 
spin parameter $a/M$ from a radio pulsar in the case of a Kerr BH with 
$a/M = 0.7$ are shown in Fig.~\ref{fig11}. The left panel is for an observer's 
viewing angle $i=90^\circ$, while the right panel for $i = 45^\circ$. 
For a Kerr BH with $a/M = 0.7$, the Hioki-Maeda distortion parameter is
$\delta = 0.0668$ ($i=90^\circ$) and $0.0371$ ($i=45^\circ$). 
The red dashed-dotted line in Fig.~\ref{fig11} indicates the objects with
the same Hioki-Maeda distortion parameter, and the two blue dashed
lines on the two sides are the boundary of the allowed region assuming 
an uncertainty on $\delta$ of 20\%\footnote{In the case $i = 90^\circ$,
the plot shows just one blue dashed line for $R/M$ because there are
no BHs with $R/M$ 1\% larger than the one of Kerr BHs.}. 
In the same way, the shadow radius
for a similar BH is $R/M = 5.20$ ($i=90^\circ$) and $R/M=5.12$ ($i=45^\circ$),
the red dashed-dotted line is the central value, while the two blue dashed
lines are the boundary of the allowed region assuming an uncertainty on 
$R/M$ of 1\%. In the case of the spin parameter $a/M$ inferred from a
radio pulsar, the uncertainty is assumed to be 1\%.

\section{Summary and conclusions \label{s-c}}

Within the next decade, VLBI facilities at mm/sub-mm wavelength will be able 
to directly image the accretion flow around SgrA$^*$, the super-massive BH 
candidate at the center of our Galaxy, and open a new window to test gravity 
in the strong field regime. In particular, it will be possible to obseve the BH 
``shadow'', whose boundary corresponds to the apparent image of the photon 
capture sphere and it is therefore determined by the spacetime geometry 
around the compact object.

At first approximation, the shadow of a BH is a circle, and its radius depend 
on the BH mass, distance, and also on the background metric. The first order 
correction to the circle is due to the BH spin, because the photon capture 
radius is larger for photons with angular momentum antiparallel to the BH 
spin (the gravitational force is stronger), and smaller in the opposite case 
(the gravitational force is weaker). The final result is that the shadow shows
a dent on one side. The magnitude of this dent can be measured in terms 
of the Hioki-Maeda distortion parameter $\delta$~\cite{maeda}. The 
measurement of $\delta$ can be used to infer one parameter of the 
background geometry. If the compact object is a Kerr BH and we have an 
independent estimate of the inclination angle $i$, $\delta$ depends only on 
the BH spin parameter, and therefore its measurement can be used to infer 
$a/M$. If we want to test the Kerr BH hypothesis, we need to measure another 
parameter of the shadow in order to break the degeneracy between the spin
and possible deviations from the Kerr solution.

In this work, we have focused the attention on the Bardeen metric, which 
is characterized by the Bardeen charge $g$ and reduces to the Kerr solution 
for $g=0$. The Bardeen solution can be seen as the prototype of a large class
of non-Kerr BH metrics, in which the line element in Boyer-Lindquist coordinates
is the same as the Kerr case with the mass $M$ replaced by a function of the 
radial coordinate $m$, which reduces to $M$ at large radii. 
We have investigated how it is possible to constrain the value
of $g/M$ from the observation of the BH shadow. We have explored three 
possibilities: the introduction of another distortion parameter of the shadow, 
$\epsilon$, the determination of the center of the shadow with respect to the 
actual position of the BH, and the estimate of the shadow radius. While all
the three approach are at least challenging, the third one may be the most
promising in the case of significant improvements of the measurement of
the mass of SgrA$^*$ and of our distance from the galactic center. Since
that can be achieved by discovering a radio pulsar with an orbital period
of a few months around SgrA$^*$, we have also briefly discussed the 
combination of the measurements of the shadow and of the orbital motion
of a similar pulsar. Such a synergy could turn out a quite interesting tool
to test the nature of the super-massive BH candidate at the center of our
Galaxy, because the shadow is sensitive to the strong gravitational field
very close to the compact object, while the pulsar can accurately probe 
the weak field at relatively large distances.

While our calculations have been done in the specific case of the Bardeen 
background, we stress that it is straightforward to repeat our study for any 
BH spacetime in which there is the Carter constant and the photon equations 
of motion are separable. The main conclusion of this work are valid even 
for those BHs.


\begin{acknowledgments}
We thank Tomohiro Harada for useful comments and suggestions.
This work was supported by the NSFC grant No.~11305038, 
the Shanghai Municipal Education Commission grant for Innovative 
Programs No.~14ZZ001, the Thousand Young Talents Program, 
and Fudan University.
\end{acknowledgments}


\end{document}